\documentclass{elsart}
\usepackage{graphicx}
\usepackage{epsfig}

\begin{document}
\begin{frontmatter}

\vspace*{-3\baselineskip}
\begin{flushleft}
 \resizebox{!}{3cm}{\includegraphics{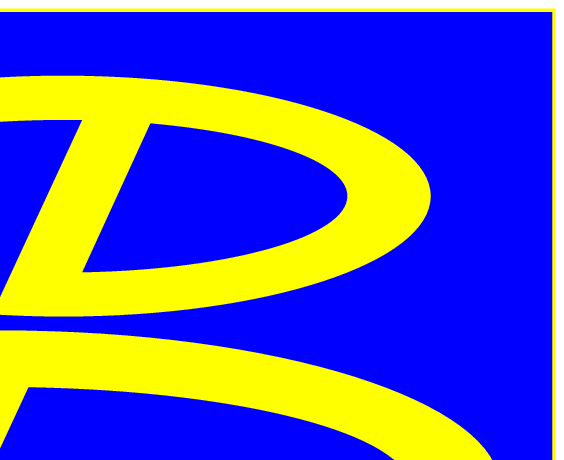}}
\end{flushleft}
\vspace*{-3cm}
\begin{flushright}
\begin{tabular}{@{} l @{}}
 Belle Preprint 2005-5\\
 KEK Preprint 2004-101\\
\end{tabular} 
\end{flushright}
\vspace*{2cm}

\title{
Measurement of $\gamma\gamma\to p\overline{p}$ production at Belle
}
\collab{Belle Collaboration}
   \author[NCU]{C.~C.~Kuo}, 
   \author[KEK]{K.~Abe}, 
   \author[KEK]{I.~Adachi}, 
   \author[Tokyo]{H.~Aihara}, 
   \author[Tsukuba]{Y.~Asano}, 
   \author[BINP]{V.~Aulchenko}, 
   \author[ITEP]{T.~Aushev}, 
   \author[Cincinnati]{S.~Bahinipati}, 
   \author[Sydney]{A.~M.~Bakich}, 
   \author[Lausanne]{A.~Bay}, 
% ...

\address[BINP]{Budker Institute of Nuclear Physics, Novosibirsk, Russia}
% ...
\address[Cincinnati]{University of Cincinnati, Cincinnati, OH, USA}
% ...
\address[KEK]{High Energy Accelerator Research Organization (KEK), Tsukuba, Japan}
% ...
\address[ITEP]{Institute for Theoretical and Experimental Physics, Moscow, Russia}
% ...
\address[Lausanne]{Swiss Federal Institute of Technology of Lausanne, EPFL, Lausanne, Switzerland}
% ...
\address[NCU]{National Central University, Chung-li, Taiwan}
% ...
\address[Sydney]{University of Sydney, Sydney, NSW, Australia}
% ...
\address[Tokyo]{Department of Physics, University of Tokyo, Tokyo, Japan}
% ...
\address[Tsukuba]{University of Tsukuba, Tsukuba, Japan}
% ...
\thanks[NovaGorica]{on leave from Nova Gorica Polytechnic, Nova Gorica, Slovenia}

\begin{abstract}
\quad A high precision study of the process $\gamma\gamma\to p\overline{p}$
has been performed using a data sample of 89~fb$^{-1}$ collected with
the Belle detector at the KEKB $e^+e^-$ collider. The cross section of
$p\overline{p}$ production has been measured at two-photon
center-of-mass (c.m.) energies between 2.025 and 4.0 GeV and in the
c.m. angular range of $|\cos{\theta^*}|<0.6$. Production of
$\gamma\gamma\to\eta_c\to p\overline{p}$ is observed and the product
of the two-photon width of the $\eta_c$ and its branching ratio to
$p\overline{p}$ is determined.

\hspace{0.1cm}
\end{abstract}


\end{frontmatter}

\newpage
\section{Introduction}

Two-photon collisions provide a clean environment for baryon pair 
production 
and such events can be produced in great abundance at a high luminosity
electron-positron collider.
Accurate measurements of such processes, in particular
$\gamma\gamma\to p\overline{p}$, is important to test existing
theoretical predictions.

General theories of hard exclusive processes 
in QCD~\cite{cz1,BL}
(see also \cite{CZR} for a review) predict that the differential 
cross section for $\gamma\gamma\to h_1h_2$
at large energies and  fixed c.m. angle ($\theta^*$) has the form
\begin{equation}
d\sigma/dt \propto s^{2-n_c} f(\theta^*) \hspace{2cm} 
(\mathrm{as}\hspace{0.2cm} s\to\infty).
\end{equation}    
Here $n_c$ is the number of elementary constituents participating in
the hard interaction, $f(\theta^*)$ is a specific function expressed
via definite integrals over hadron wave functions, $s$ is the square
of the c.m. energy of the two-photon system, and $t$ is the square of
the four-momentum transfer from a photon to hadron.  The first
estimate of the cross section for $\gamma\gamma\to p\overline{p}$ was
obtained in the three-quark picture ($n_c=8$) \cite{FA,FA2}, using the
proton wave function based on QCD sum rules \cite{CZ}.  Previous
measurements \cite{TA,CL,VN,OP,L3} in the
$W_{\gamma\gamma}(\equiv\sqrt{s})$ range between 2.5 and 3.0 GeV gave
cross sections one order of magnitude larger than this expectation.
To explain these experimental observations, various model-dependent
approaches were suggested. For example, in the diquark model
\cite{MA,KP,CF5} the proton is considered to be a quark-diquark
system.  In this case $n_c=6$ and a diquark form factor is introduced,
so that Eq.~(1) becomes $ d\sigma/dt \propto s^{-4} |F|^2 $, where $F$
may depend on $s$. Asymptotically, $F\to f(\theta^*)/s$ \cite{MA}, and
the behavior $d\sigma/dt \propto s^{-6}$ is recovered.  These results
exhibit better agreement with measurements of the absolute size of the
cross section for $W_{\gamma\gamma}=2.5$ - 3.0 GeV.

Other approaches have been developed recently.
The handbag model~\cite{HB} has been developed for large momentum 
transfers, and the calculations have been applied at medium energies   
($W_{\gamma\gamma}>2.55$ GeV) with large uncertainty bands.
In Ref.~\cite{OD}, the Veneziano model is applied in an unmodified form to 
the process, and fair agreement with data is obtained without adjustable 
parameters.

Recently, the measured energy range for $\gamma\gamma\to p\overline{p}$
has been extended to $W_{\gamma\gamma}=4$ GeV and above \cite{OP,L3},
but with very limited statistics for $W_{\gamma\gamma}>3$ GeV. 
Furthermore, $p\overline{p}\to\gamma\gamma$ experiments
give the cross section for the inverse process 
at $W_{\gamma\gamma}=3.2$ - $3.7$ 
GeV \cite{Flab}. To test 
QCD predictions, it is very important 
to improve the statistics at higher energies. 
Moreover, an accurate measurement with higher statistics 
for $\gamma\gamma\to p\overline{p}$
is crucial in the study of the interactions involved. 
This paper presents the Belle measurement of the 
$\gamma\gamma\to p\overline{p}$ 
cross section for $W_{\gamma\gamma}$ between 2.025 to 4.0 GeV
and $|\cos{\theta^*}|<0.6$, 
using a data sample
corresponding to an integrated luminosity of 89~fb$^{-1}$.   


\section{Experimental apparatus and event selection}

Experimental data are recorded with the Belle detector \cite{belle} at
KEKB \cite{kekb}, which is an asymmetric $e^+e^-$ collider running at
10.58~GeV c.m. energy.  In the laboratory frame, the direction of the
positron beam is taken to define the $-z$ direction.  For the analyses
in this paper, the following Belle subsystems are of importance: the
central drift chamber (CDC), the aerogel Cherenkov counters (ACC), the
time-of-flight scintillation counters (TOF) and the CsI
electromagnetic calorimeter (ECL), all of which are located in a 1.5 T
solenoidal magnetic field. 
The CDC measures the momenta of charged particles and provides particle
identification information by precise ($6\%$) $dE/dx$ measurement,
allowing separation of 
protons from other particles for momentum up to 1 GeV/$c$. 
The TOF measures the time of flight of particles 
with a 0.1 ns timing resolution, which is powerful for $p/K$ 
separation for momentum up to 2 GeV/$c$. 
The ECL detects photons and is used to
reject electrons by measuring the deposited energy with a
resolution of $\sigma_E/E=1.5\%$ $(2.0\%)$ at 1 GeV (0.1 GeV).  
Using the number of photoelectrons observed, the ACC extends particle
identification beyond that of the CDC $dE/dx$ and TOF and is effective 
in the suppression of highly relativistic $\pi^{\pm},\mu^{\pm}$ and $e^{\pm}$ 
up to a momentum of 3.5 GeV/$c$.  

Through the process $e^+e^-\to e^+e^-\gamma^*\gamma^*\to e^+e^- p\overline{p}$,
exclusive $p\overline{p}$ pairs are produced in quasi-real two-photon 
collisions,
where the scattered $e^+$ and $e^-$ are lost down the beam-pipe, 
and only the $p$ and $\overline{p}$ can be detected. The $\gamma\gamma$
axis thus approximates the beam direction in the 
$e^+e^-$ c.m. frame.   
Candidate events are searched for in a data stream 
where the sum of the magnitudes of momentum of all charged tracks and 
the total ECL energy are restricted to 6 GeV/$c$ and 6 GeV, respectively. 
Events are required to have exactly two tracks
of opposite charge satisfying the following conditions in the laboratory 
frame: 
$p_t>0.35$ GeV/$c$,
$dr<1$ cm and $|dz|<5$ cm.  Here $p_t$ is the transverse
momentum and $dr$ and $dz$ are the radial and axial coordinates of the 
point of 
closest approach to the nominal collision point, respectively. 
Both tracks are required to hit the TOF counters. 
The invariant mass of the two tracks and the squared missing mass of the 
event, assuming the two tracks are massless, are required to be smaller 
than 4.5 GeV/$c^2$
and larger than 2 GeV$^2$/$c^4$, respectively. 
A good transverse momentum balance in the $e^+e^-$ c.m. frame 
is also required:
$|\Sigma p_t^*|\equiv 
|\vec{p}_{t_1}^{\hspace{0.1cm}*}+\vec{p}_{t_2}^{\hspace{0.1cm}*}|
<0.2$ GeV/$c$,
where $\vec{p}_{t_1}^{\hspace{0.1cm}*}$ and $\vec{p}_{t_2}^{\hspace{0.1cm}*}$
denote the transverse momenta of the two tracks 
in that frame, with respect to the $e^+e^-$ beam-axis.

The selected events are dominated by $\gamma\gamma\to e^+e^-$, 
$\mu^+\mu^-$, $\pi^+\pi^-$ and $K^+K^-$ up to this stage. 
Events with $p\overline{p}$ are separated 
from the others by a particle identification (PID) algorithm, which is applied 
to each individual track under the following conditions:
1) the difference between the measured and the expected CDC $dE/dx$
is less than 4 times the resolution: 
$\chi^2_{dE/dx}\equiv 
[\Delta (dE/dx) / \sigma_{dE/dx}]^2
<4^2$; 
2) the ratio of the associated ECL energy to the momentum 
is less than 0.9, which is only applied 
to the positively charged track;
3) the number of the photoelectrons in ACC counters associated with the 
track is less than 4, and this condition removes a large part of
high-momentum $\pi^{\pm},\mu^{\pm}$ and $e^{\pm}$; 
4) the likelihoods for each particle assignment are combined to determine  
the normalized likelihood,  
$\lambda_p\equiv L_p/(L_p+L_K+L_{\pi}+L_{\mu}+L_e)$, which
has to be larger than 0.8 (Fig.~1).
In these likelihoods, $L\equiv exp[-\frac{1}{2}(\chi^2_{dE/dx}+\chi^2_T)]$
is calculated using information from the CDC ($dE/dx$) and the TOF (time of 
flight $T$).
Here $\chi^2_T\equiv (\Delta T/\sigma_T)^2 $, $\Delta T$ is the difference 
between the measured and the expected values for $T$, and $\sigma_T$ is the 
timing resolution. 
A combined use of CDC and TOF allows $p(\overline{p})$ separation 
from other particles, in particular $K^{\pm}$, for momentum up to 2 GeV/$c$.

\begin{figure}
\begin{center}
\includegraphics[width=7.cm,height=7.cm]{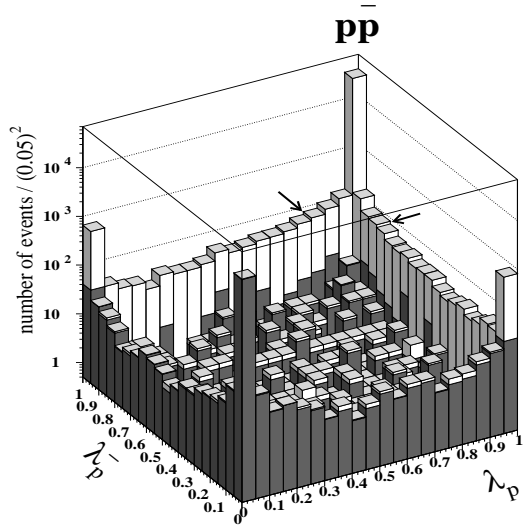}
\caption{Two-dimensional distribution of the normalized likelihood for 
$p\overline{p}$ identification, for the events passing all selection
criteria except the cuts on the normalized likelihood indicated by
the arrows. The dark parts show the events satisfying 
$\lambda_{\mathrm{x}}\equiv L_{\mathrm{x}}/(L_p+L_K+L_{\pi}+L_{\mu}+L_e)>0.01$ 
for both tracks using the same $\mathrm{x}$, where $\mathrm{x}$ is 
$K$, $\pi$, $\mu$ or $e$.}
\end{center}
\end{figure}

\begin{figure}[h]
\begin{center}
\includegraphics[width=6.8cm,height=6.8cm]{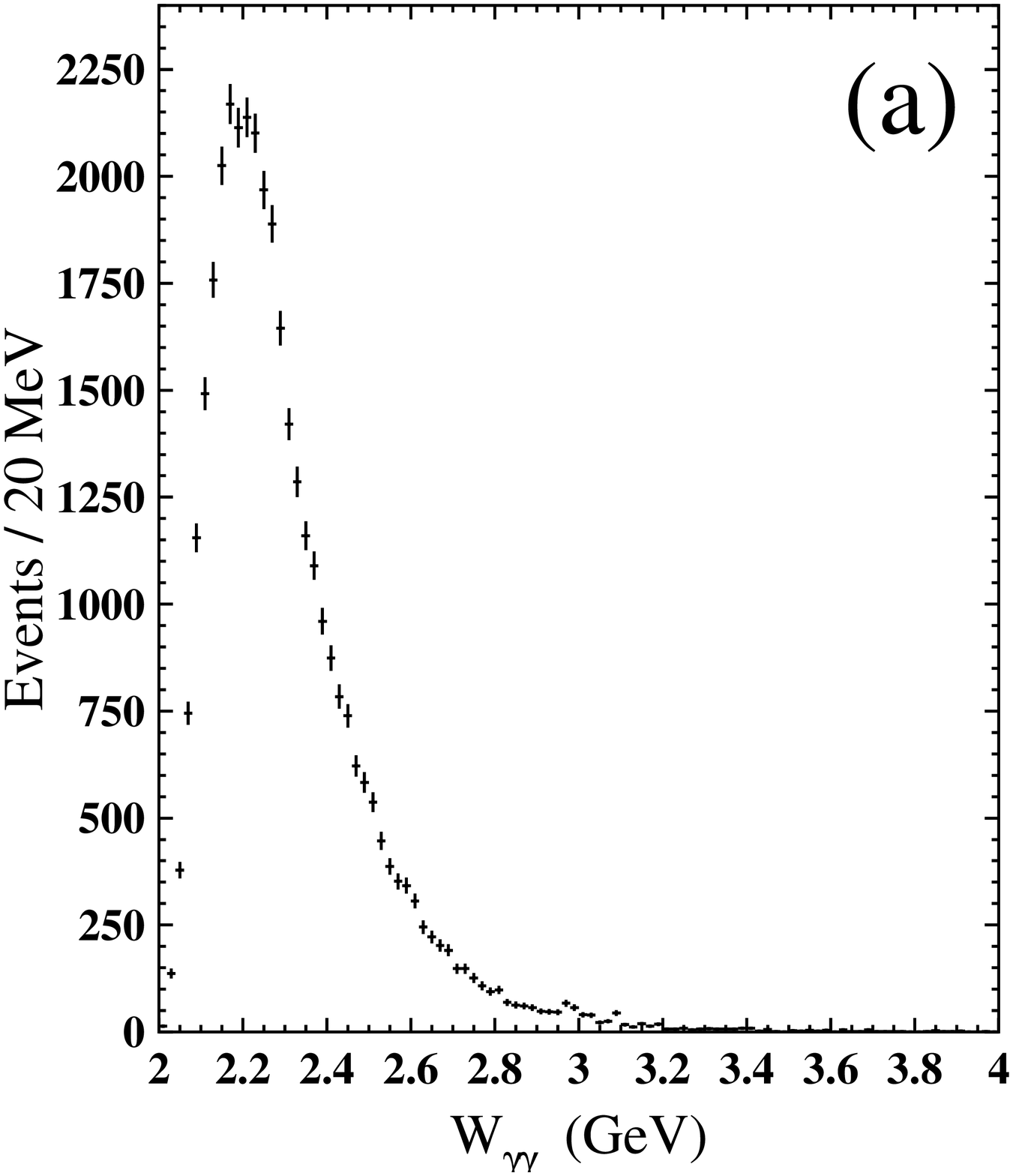}
\includegraphics[width=6.8cm,height=6.8cm]{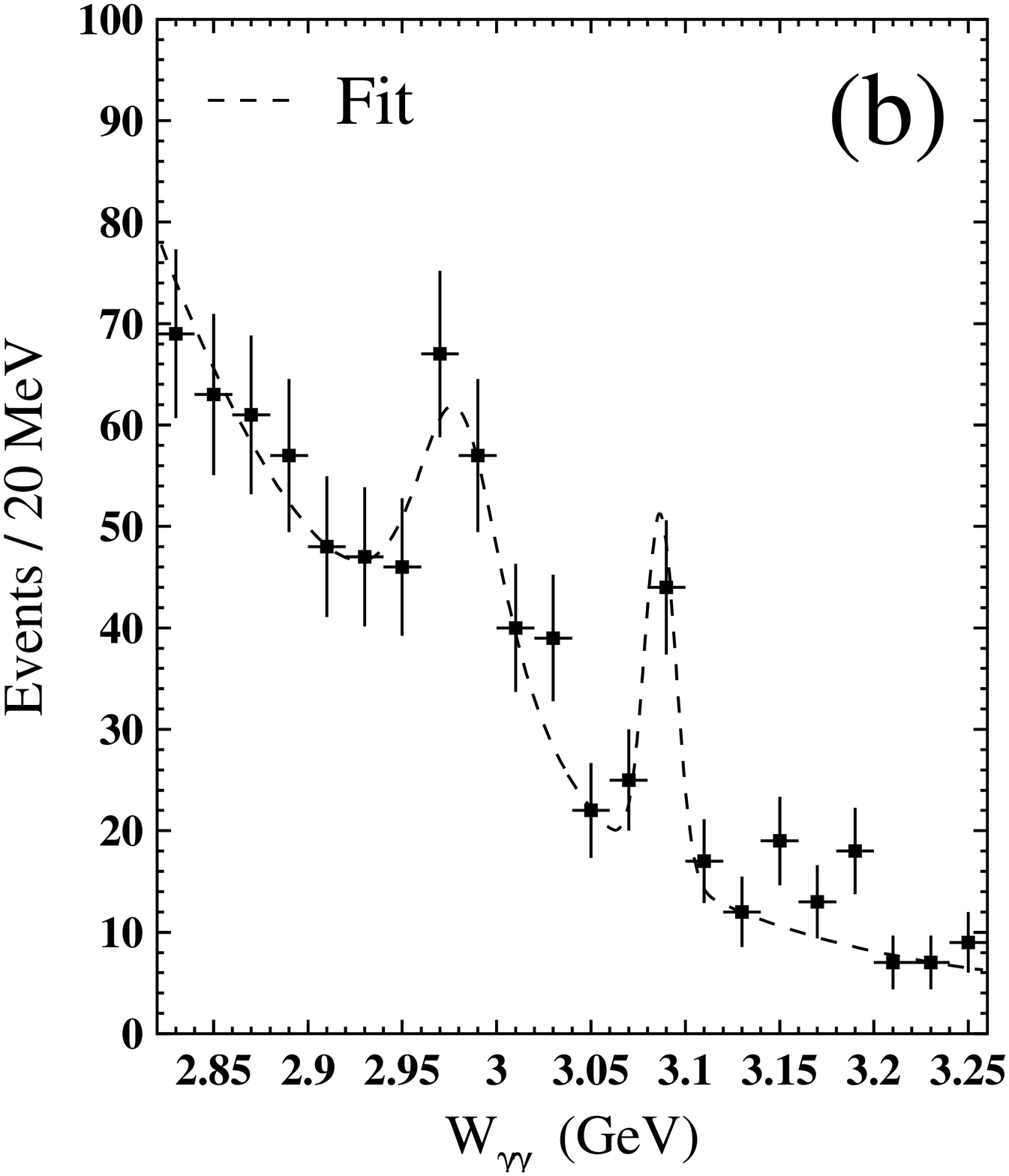}
\caption{$W_{\gamma\gamma}$ distribution of events  
passing all the selection criteria: (a) $W_{\gamma\gamma}=2$ - 4 GeV; 
(b) a close-up view of (a) for $W_{\gamma\gamma}=2.82$ - 3.26 GeV.
}
\end{center}
\end{figure}

For $W_{\gamma\gamma}=2$ - 4 GeV a
total 36094 events survive all of the selection criteria.
Their distribution in $W_{\gamma\gamma}$ is shown in Fig.~2. 
A peak around 2.98 GeV can be identified as the $\eta_c(2980)$ 
resonance \cite{PDG}.
A much narrower peak at 3.08 - 3.10 GeV, corresponding to an 
excess of $26\pm8$ events relative to the continuum,
could be attributed to backgrounds from  
radiative return to $J/\psi$,
and the enhancement
is in agreement with expectations 
based on this assumption \cite{BN}.
Fitting the data from 2.6 to 3.7 GeV
with a smooth (exponential of a fifth-order polynomial) function
for the continuum, a Breit-Wigner function for the $\eta_c$ 
and a Gaussian for the $J/\psi$,
a total $\eta_c$ yield of $156.9\pm33.3$ events is obtained.
The statistical significance of the $\eta_c$ signal is
$5.3~\sigma$, defined as $\sqrt{-2~{\rm ln}(L_0/L_{S})}$, 
where $L_S$ and $L_0$ denote
the maximum likelihoods
of the fits with and without a signal component, respectively.


\section{Monte Carlo simulation}

Monte Carlo samples for the following 
channels have been generated: $e^+e^-\to e^+e^-X$ where $X$ is 
$p\overline{p}$, $K^+K^-$, $\pi^+\pi^-$, 
$\mu^+\mu^-$, $e^+e^-$ and $p\overline{p}\pi^0$.
Hadron pair, lepton pair and $\gamma\gamma\to p\overline{p}\pi^0$
events are generated by the codes TREPS~\cite{TRE}, AAFH~\cite{AAF} and
GGLU~\cite{GGL}, respectively.
Event generation is followed by a 
detector simulation based on GEANT3 \cite{GEA} and a trigger
simulation. The selection criteria
described in Section 2 are then applied to these Monte Carlo events.    

Because the acceptance depends on both $W_{\gamma\gamma}$ and 
$|\cos{\theta^*}|$, the signal ($\gamma\gamma\to p\overline{p}$) 
efficiencies are determined for
a number of two-dimensional bins of the two variables.      
Other channels are generated for the study of residual backgrounds.
Similar to $\gamma\gamma\to p\overline{p}$, 
the selection efficiencies for $\gamma\gamma\to K^+K^-$ and
$\gamma\gamma\to p\overline{p}\pi^0$
are evaluated within each narrow bin. 
For the $\gamma\gamma\to \pi^+\pi^-$, 
$\mu^+\mu^-$ and $e^+e^-$ channels, 
realistic distributions
are generated \cite{AAF,PI1,PI2}. 
Events from those samples that survive the selection criteria are 
referred to as the expected residual backgrounds.

From Monte Carlo simulation, the overall efficiency of 
$\gamma\gamma\to p\overline{p}$ for $|\cos{\theta^*}|<0.1$  
ranges from $\sim 3\%$ at 
$W_{\gamma\gamma}=2$ GeV to $\sim 32\%$ at $W_{\gamma\gamma}=4$ GeV (Fig.~3).

\begin{figure}
\begin{center}
\includegraphics[width=7.cm,height=7.cm]{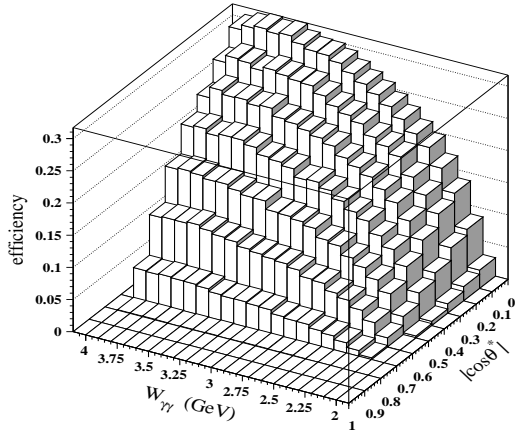}
\caption{Overall detection efficiency of $\gamma\gamma\to p\overline{p}$
as a function of $W_{\gamma\gamma}$ and $|\cos{\theta^*}|$.}
\end{center}
\end{figure}


\section{Measurement of the cross sections for 
$\gamma\gamma\to p\overline{p}$}

\begin{figure}
\begin{center}
\includegraphics[width=12.cm,height=12.cm]{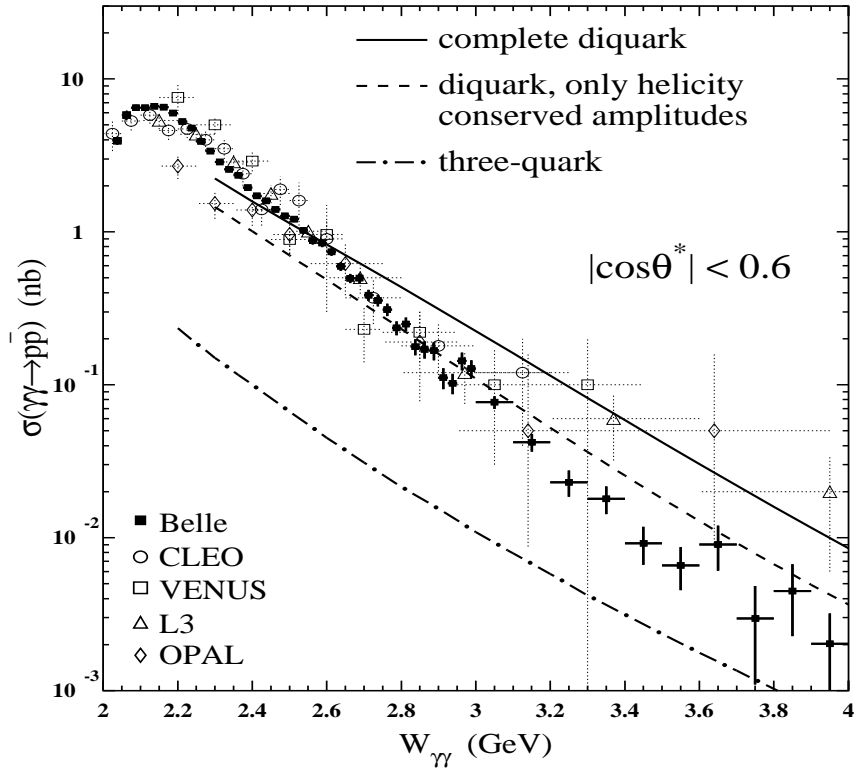}
\caption{Measured cross sections for $\gamma\gamma\to p\overline{p}$.
For the Belle, CLEO \cite{CL} and VENUS \cite{VN} results, the error bars 
are purely 
statistical; while for OPAL \cite{OP} and L3 \cite{L3}, 
both statistical and systematic 
uncertainties are included.
Theoretical prediction curves shown are from \cite{CF5} (diquark) 
and \cite{FA} (three-quark).}
\end{center}
\end{figure}

Dividing $|\cos{\theta^*}|$ into bins of 0.1, 
$W_{\gamma\gamma}$ into bins of 25 MeV (for
2 - 3 GeV) and 100 MeV (for
3 - 4 GeV),
the number of events selected from the data, 
$\Delta N(W_{\gamma\gamma},|\cos{\theta^*}|)$, 
is determined for each of the two-dimensional bins.
The efficiency $\varepsilon(W_{\gamma\gamma},|\cos{\theta^*}|)$ 
is also evaluated from Monte Carlo simulation for each bin. 
The ratio of $\Delta N$ to $\varepsilon$ is then converted to the 
differential cross section, according to the formula:
\begin{equation}
\frac{d\sigma_{\gamma\gamma\to p\overline{p}}
(W_{\gamma\gamma})}{d|\cos{\theta^*}|}=
\frac{\Delta N (1-f)/\varepsilon}
{L_{\rm int}\hspace{0.15cm}
\frac{dL_{\gamma\gamma}}{dW_{\gamma\gamma}}\hspace{0.15cm}
\Delta W_{\gamma\gamma}\hspace{0.15cm} 
\Delta |\cos{\theta^*}|}
\end{equation}      
where $f$ is the fraction of residual background in the data,
$L_{\rm int}$ is the integrated luminosity and 
$\frac{dL_{\gamma\gamma}}{dW_{\gamma\gamma}}$ is the luminosity function.
Here $L_{\rm int}=88.96$~fb$^{-1}$, with a systematic uncertainty of $1.4\%$.
The luminosity function $\frac{dL_{\gamma\gamma}}{dW_{\gamma\gamma}}$,
as a function of $W_{\gamma\gamma}$,
is defined by
\begin{equation}
\sigma_{e^+e^-\to e^+e^- p\overline{p}}=
\int \sigma_{\gamma\gamma\to p\overline{p}}(W_{\gamma\gamma})\hspace{0.25cm}
\frac{dL_{\gamma\gamma}(W_{\gamma\gamma})}{dW_{\gamma\gamma}}\hspace{0.25cm}
dW_{\gamma\gamma} 
\end{equation}
and is calculated by TREPS \cite{TRE} using the equivalent photon
approximation method \cite{BN}.  For the calculation of the luminosity
function, the effects from longitudinal photons are neglected. 
For simulation in TREPS, the
maximum virtuality of each of the two photons, $Q_1^2$ and $Q_2^2$, is
limited to 1 GeV$^2$. Moreover, a form factor term is introduced for
the high-$Q^2$ suppression effect,
$(1+Q_1^2/W_{\gamma\gamma}^2)^{-2}(1+Q_2^2/W_{\gamma\gamma}^2)^{-2}$.
The systematic uncertainty of the luminosity function is estimated by
comparing the kinematic distributions of the two-photon system for the
events generated with TREPS to those from a QED calculation that
includes all order $\alpha^4$ diagrams \cite{AAF}. Within the range
$W_{\gamma\gamma}=2$ - 4 GeV agreement within $3$ - $5\%$ 
was reported \cite{TRE,LFE}.

\begin{table}[ht]
\begin{center}
{\footnotesize
\begin{tabular}{|c|c||c|l|}
\hline
 $W_{\gamma\gamma}$ (GeV) & 
 $\sigma(\gamma\gamma\to p\overline{p})$ (nb) &
 $W_{\gamma\gamma}$ (GeV) & 
 \hspace{0.7cm}$\sigma(\gamma\gamma\to p\overline{p})$ (nb) \\
\hline\hline 
  2.025 - 2.050  &  3.95 $\pm$ 0.25 $\pm$ 0.28 & 2.650 - 2.675  &  0.50\hspace{0.37cm} $\pm$ 0.03\hspace{0.38cm} $\pm$ 0.05\\   
  2.050 - 2.075  &  5.79 $\pm$ 0.40 $\pm$ 0.41 & 2.675 - 2.700  &  0.50\hspace{0.37cm} $\pm$ 0.04\hspace{0.38cm} $\pm$ 0.05\\   
  2.075 - 2.100  &  6.48 $\pm$ 0.29 $\pm$ 0.46 & 2.700 - 2.725  &  0.39\hspace{0.37cm} $\pm$ 0.03\hspace{0.38cm} $\pm$ 0.04\\   
  2.100 - 2.125  &  6.49 $\pm$ 0.17 $\pm$ 0.49 & 2.725 - 2.750  &  0.36\hspace{0.37cm} $\pm$ 0.03\hspace{0.38cm} $\pm$ 0.03\\   
  2.125 - 2.150  &  6.64 $\pm$ 0.16 $\pm$ 0.50 & 2.750 - 2.775  &  0.31\hspace{0.37cm} $\pm$ 0.03\hspace{0.38cm} $\pm$ 0.03\\   
  2.150 - 2.175  &  6.54 $\pm$ 0.14 $\pm$ 0.50 & 2.775 - 2.800  &  0.24\hspace{0.37cm} $\pm$ 0.02\hspace{0.38cm} $\pm$ 0.02\\   
  2.175 - 2.200  &  5.97 $\pm$ 0.13 $\pm$ 0.47 & 2.800 - 2.825  &  0.25\hspace{0.37cm} $\pm$ 0.03\hspace{0.38cm} $\pm$ 0.02\\   
  2.200 - 2.225  &  5.26 $\pm$ 0.11 $\pm$ 0.42 & 2.825 - 2.850  &  0.18\hspace{0.37cm} $\pm$ 0.02\hspace{0.38cm} $\pm$ 0.02\\   
  2.225 - 2.250  &  4.77 $\pm$ 0.10 $\pm$ 0.39 & 2.850 - 2.875  &  0.17\hspace{0.37cm} $\pm$ 0.02\hspace{0.38cm} $\pm$ 0.02\\   
  2.250 - 2.275  &  3.91 $\pm$ 0.09 $\pm$ 0.32 & 2.875 - 2.900  &  0.17\hspace{0.37cm} $\pm$ 0.02\hspace{0.38cm} $\pm$ 0.02\\   
  2.275 - 2.300  &  3.38 $\pm$ 0.08 $\pm$ 0.28 & 2.900 - 2.925  &  0.111\hspace{0.18cm} $\pm$ 0.017\hspace{0.195cm} $\pm$ 0.010\\   
  2.300 - 2.325  &  2.86 $\pm$ 0.07 $\pm$ 0.24 & 2.925 - 2.950  &  0.101\hspace{0.18cm} $\pm$ 0.016\hspace{0.195cm} $\pm$ 0.009\\   
  2.325 - 2.350  &  2.56 $\pm$ 0.07 $\pm$ 0.22 & 2.950 - 2.975  &  0.143\hspace{0.18cm} $\pm$ 0.019\hspace{0.195cm} $\pm$ 0.013\\   
  2.350 - 2.375  &  2.34 $\pm$ 0.07 $\pm$ 0.21 & 2.975 - 3.000  &  0.128\hspace{0.18cm} $\pm$ 0.017\hspace{0.195cm} $\pm$ 0.011\\   
  2.375 - 2.400  &  1.94 $\pm$ 0.06 $\pm$ 0.17 & 3.000 - 3.100  &  0.077\hspace{0.18cm} $\pm$ 0.007\hspace{0.195cm} $\pm$ 0.009\\   
  2.400 - 2.425  &  1.72 $\pm$ 0.06 $\pm$ 0.16 & 3.100 - 3.200  &  0.042\hspace{0.18cm} $\pm$ 0.006\hspace{0.195cm} $\pm$ 0.004\\   
  2.425 - 2.450  &  1.59 $\pm$ 0.05 $\pm$ 0.15 & 3.200 - 3.300  &  0.023\hspace{0.18cm} $\pm$ 0.005\hspace{0.195cm} $\pm$ 0.002\\   
  2.450 - 2.475  &  1.40 $\pm$ 0.05 $\pm$ 0.13 & 3.300 - 3.400  &  0.018\hspace{0.18cm} $\pm$ 0.004\hspace{0.195cm} $\pm$ 0.002\\   
  2.475 - 2.500  &  1.27 $\pm$ 0.05 $\pm$ 0.12 & 3.400 - 3.500  &  0.0092 $\pm$ 0.0026 $\pm$ 0.0009\\
  2.500 - 2.525  &  1.21 $\pm$ 0.05 $\pm$ 0.11 & 3.500 - 3.600  &  0.0066 $\pm$ 0.0021 $\pm$ 0.0007\\
  2.525 - 2.550  &  1.03 $\pm$ 0.05 $\pm$ 0.10 & 3.600 - 3.700  &  0.0090 $\pm$ 0.0030 $\pm$ 0.0010\\
  2.550 - 2.575  &  0.88 $\pm$ 0.05 $\pm$ 0.08 & 3.700 - 3.800  &  0.0030 $\pm$ 0.0019 $\pm$ 0.0003\\
  2.575 - 2.600  &  0.84 $\pm$ 0.05 $\pm$ 0.08 & 3.800 - 3.900  &  0.0045 $\pm$ 0.0022 $\pm$ 0.0005\\
  2.600 - 2.625  &  0.74 $\pm$ 0.04 $\pm$ 0.07 & 3.900 - 4.000  &  0.0020 $\pm$ 0.0012 $\pm$ 0.0003\\
  2.625 - 2.650  &  0.60 $\pm$ 0.04 $\pm$ 0.06 & & \\  
\hline
\end{tabular}
}
\vspace{0.25cm}
\caption{Measured cross sections for 
$\gamma\gamma\to p\overline{p}$ ($|\cos{\theta^*}|<0.6$).
The first error is statistical and the second is systematic.}
\end{center}
\end{table}

The cross section $\sigma_{\gamma\gamma\to p\overline{p}}(W_{\gamma\gamma})$
is obtained by a summation over $|\cos{\theta^*}|$:  
$\sum$ 
$(d\sigma_{\gamma\gamma\to p\overline{p}}(W_{\gamma\gamma})/
d|\cos{\theta^*}|)$  
$\Delta |\cos{\theta^*}|$, 
with the restriction of $|\cos{\theta^*}|<0.6$, due to polar angular 
coverage limits 
of the TOF system.
The results are summarized in Table 1 and Fig.~4.
The contribution from $\gamma\gamma\to\eta_c\to p\overline{p}$ is 
included in these results.   
For the two cross section measurements in the lowest $W_{\gamma\gamma}$
bins (2.025 - 2.075 GeV),
the efficiencies are extremely small at larger $|\cos{\theta^*}|$ and 
data are only available up to $|\cos{\theta^*}|=0.4$ and 0.5, respectively.
We thus fit a second-order polynomial function of $\cos^2{\theta^*}$ to
these differential cross sections 
and arrive at a result by integrating the fit over $|\cos{\theta^*}|$ 
up to 0.6.
For comparison, we also show in Fig.~4 the results from 
previous measurements \cite{CL,VN,OP,L3}.

Fig.~5 shows the angular dependence of the differential cross sections  
measured in 11 ranges of $W_{\gamma\gamma}$ separately. The $\eta_c$ region
(2.9 - 3.1 GeV) is skipped. For $W_{\gamma\gamma}<2.4$ GeV,
the differential cross sections decrease as $|\cos{\theta^*}|$ increases;
while for $W_{\gamma\gamma}>2.6$ GeV, the opposite trend is observed.
The transition occurs around $W_{\gamma\gamma}=2.5$ GeV. 
Similar results are shown in Fig.~6 for three larger ranges of 
$W_{\gamma\gamma}$ and are summarized in Table 2. For comparison, 
previous measurements \cite{CL,OP,L3}
are also shown in Fig.~6.
Further discussion is given in Section 6.

\begin{figure}
\begin{center}
\includegraphics[width=4.55cm,height=4.55cm]{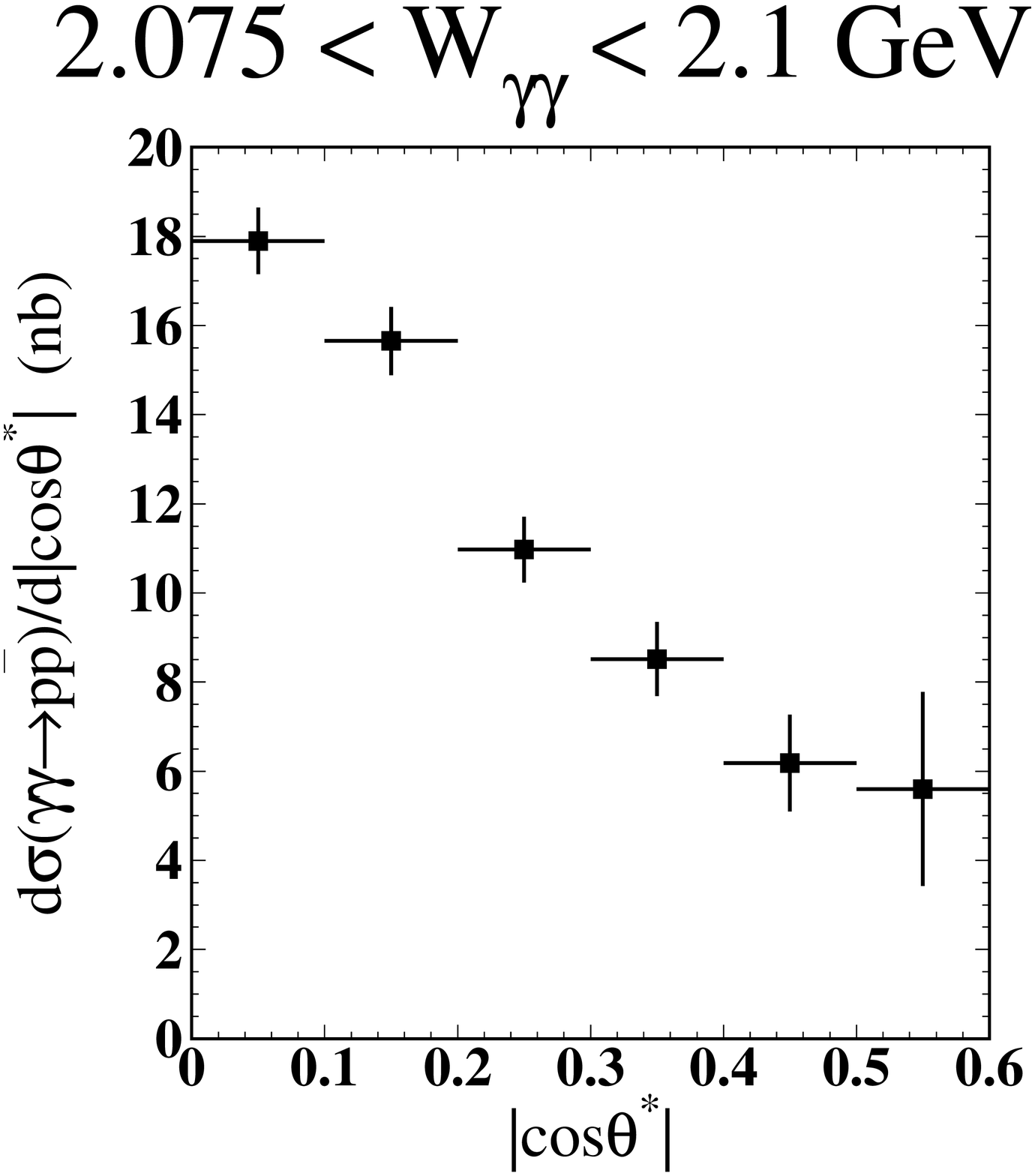}
\includegraphics[width=4.55cm,height=4.55cm]{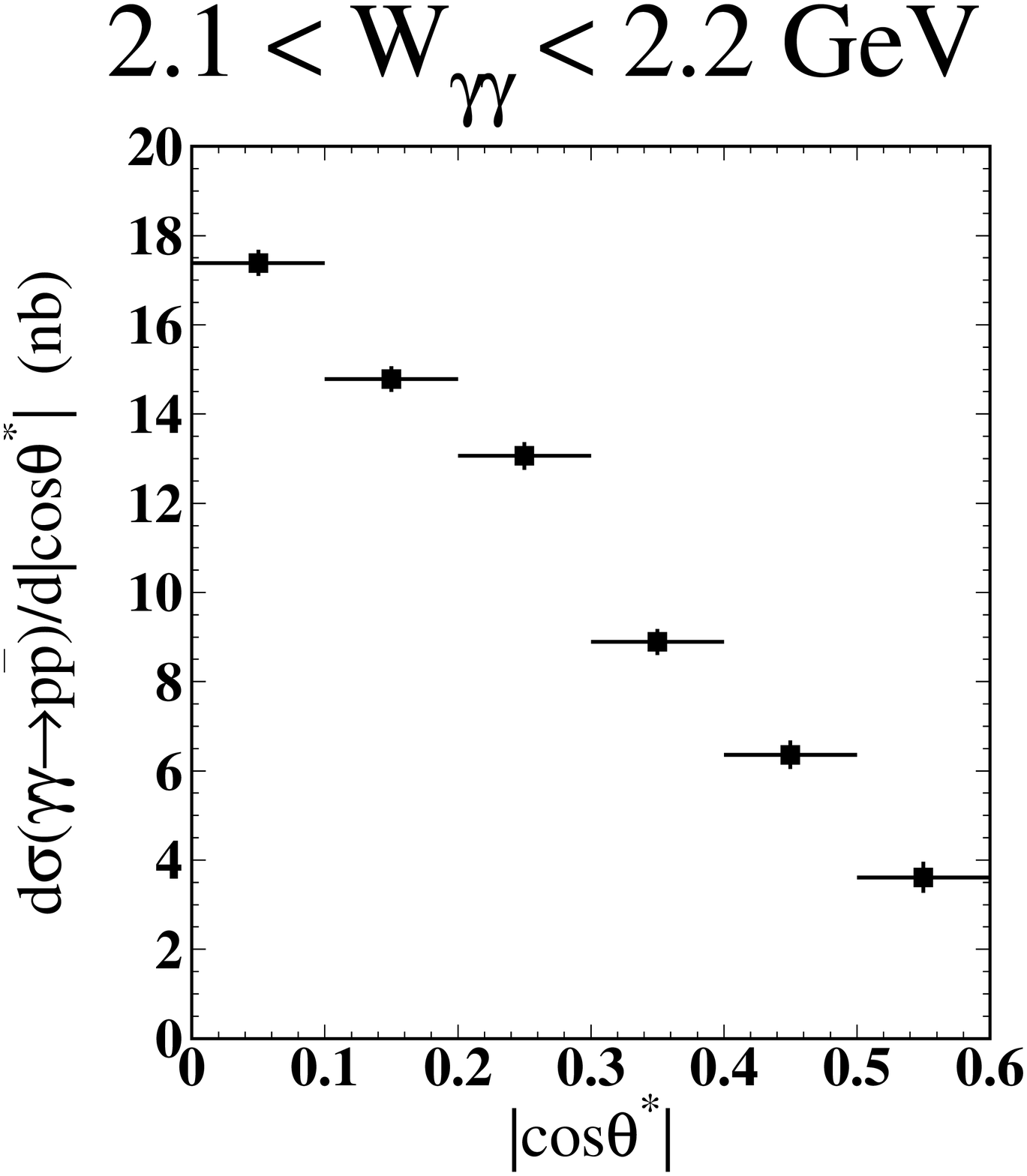}
\includegraphics[width=4.55cm,height=4.55cm]{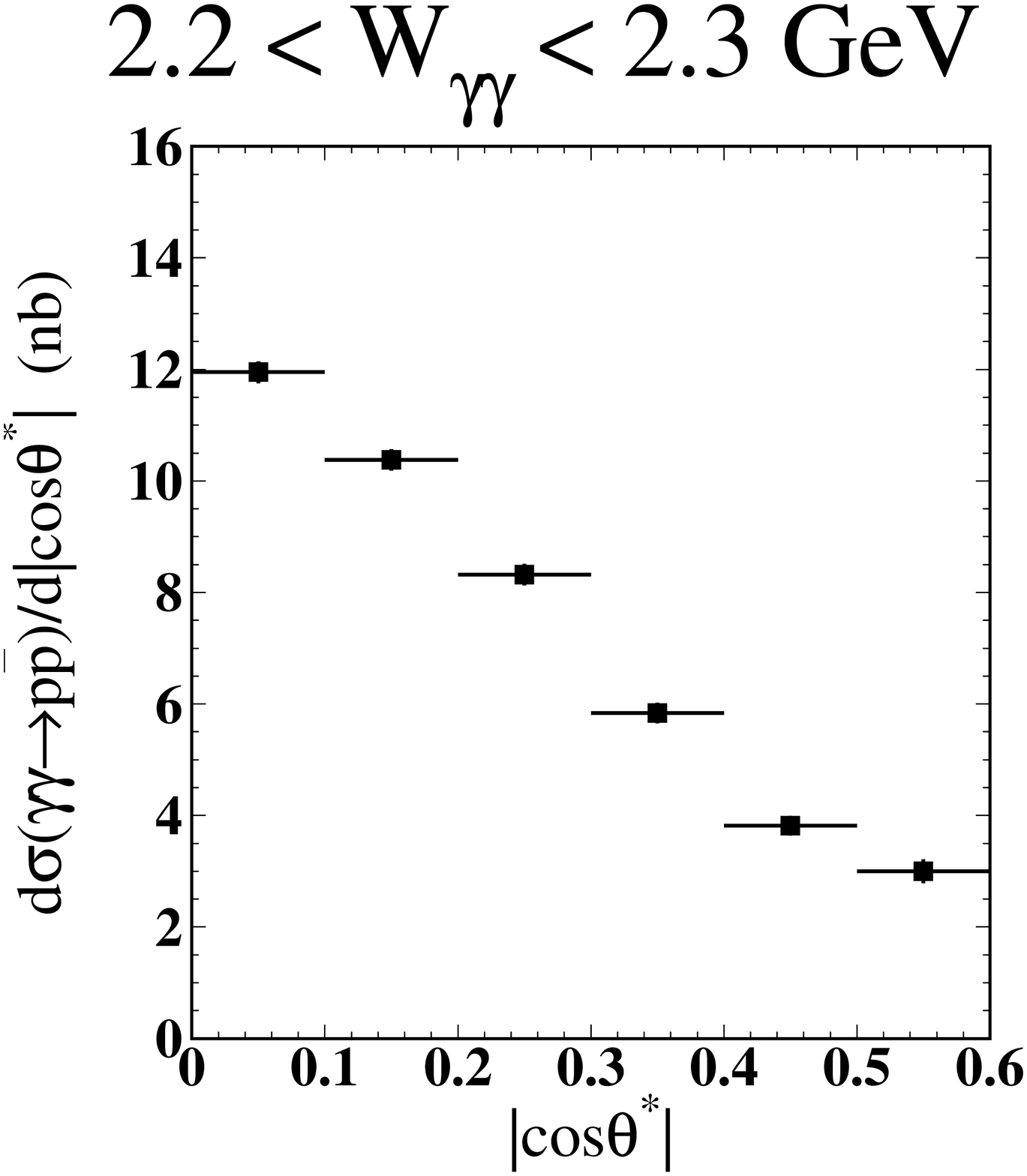}
\includegraphics[width=4.55cm,height=4.55cm]{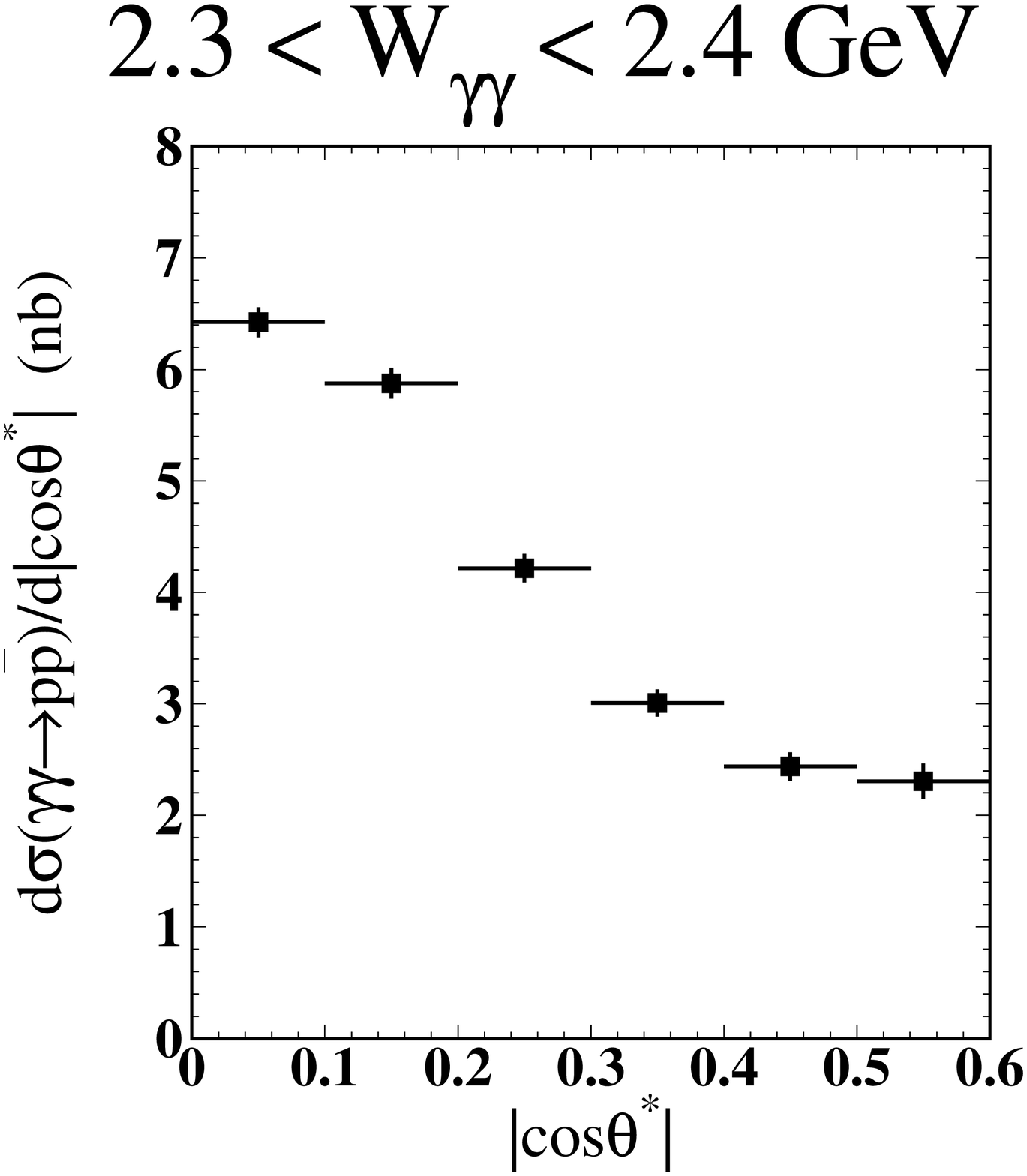}
\includegraphics[width=4.55cm,height=4.55cm]{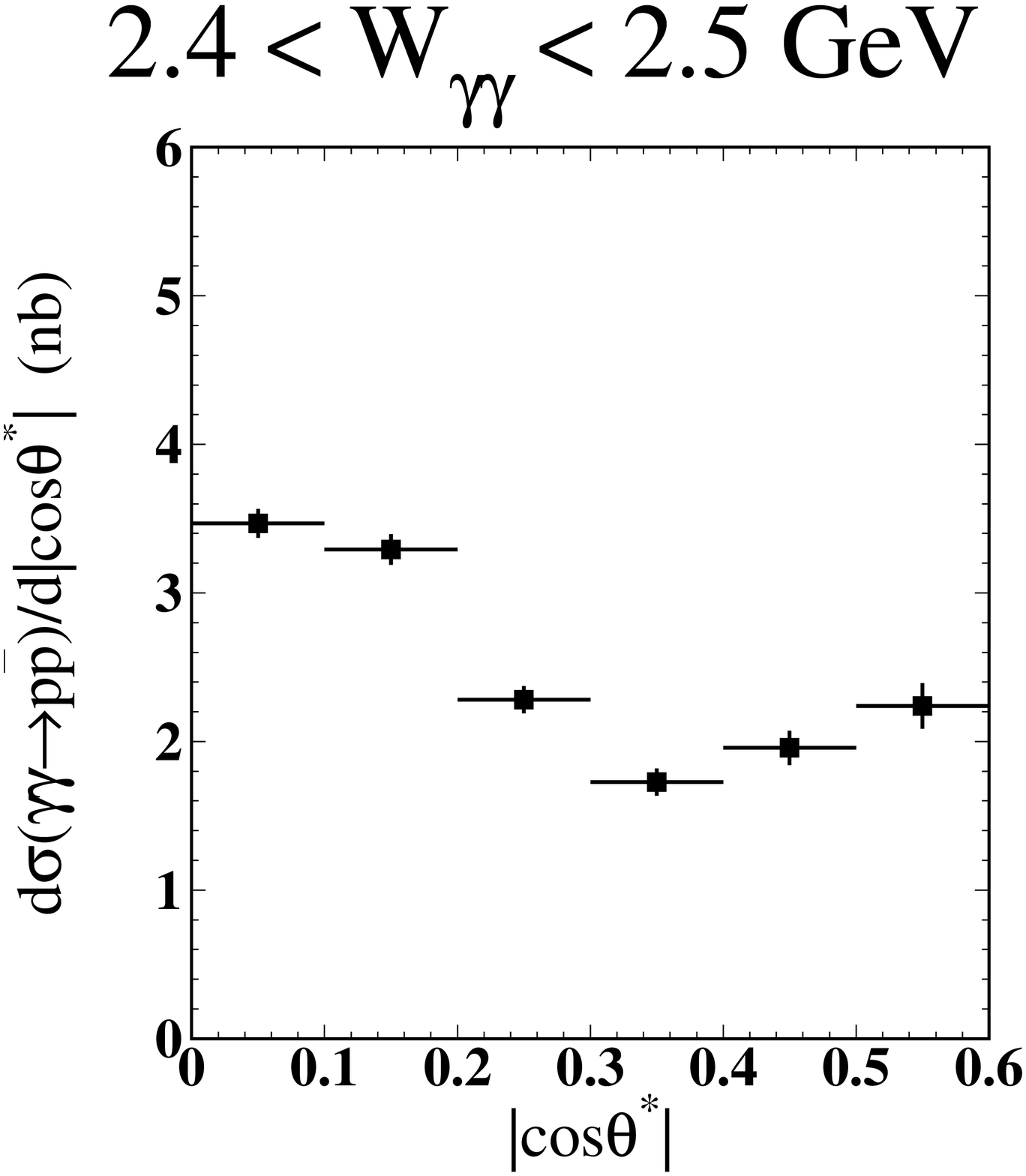}
\includegraphics[width=4.55cm,height=4.55cm]{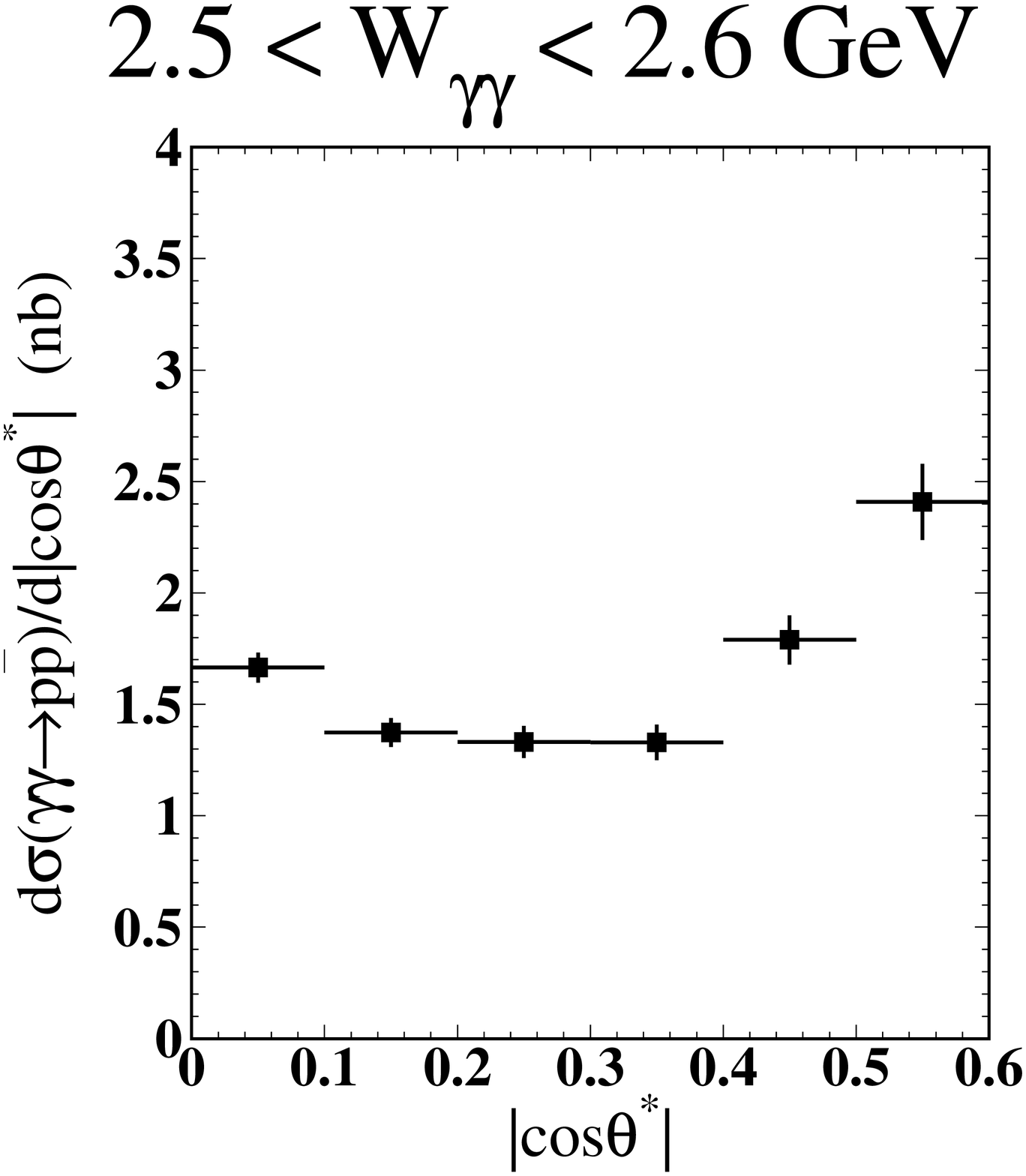}
\includegraphics[width=4.55cm,height=4.55cm]{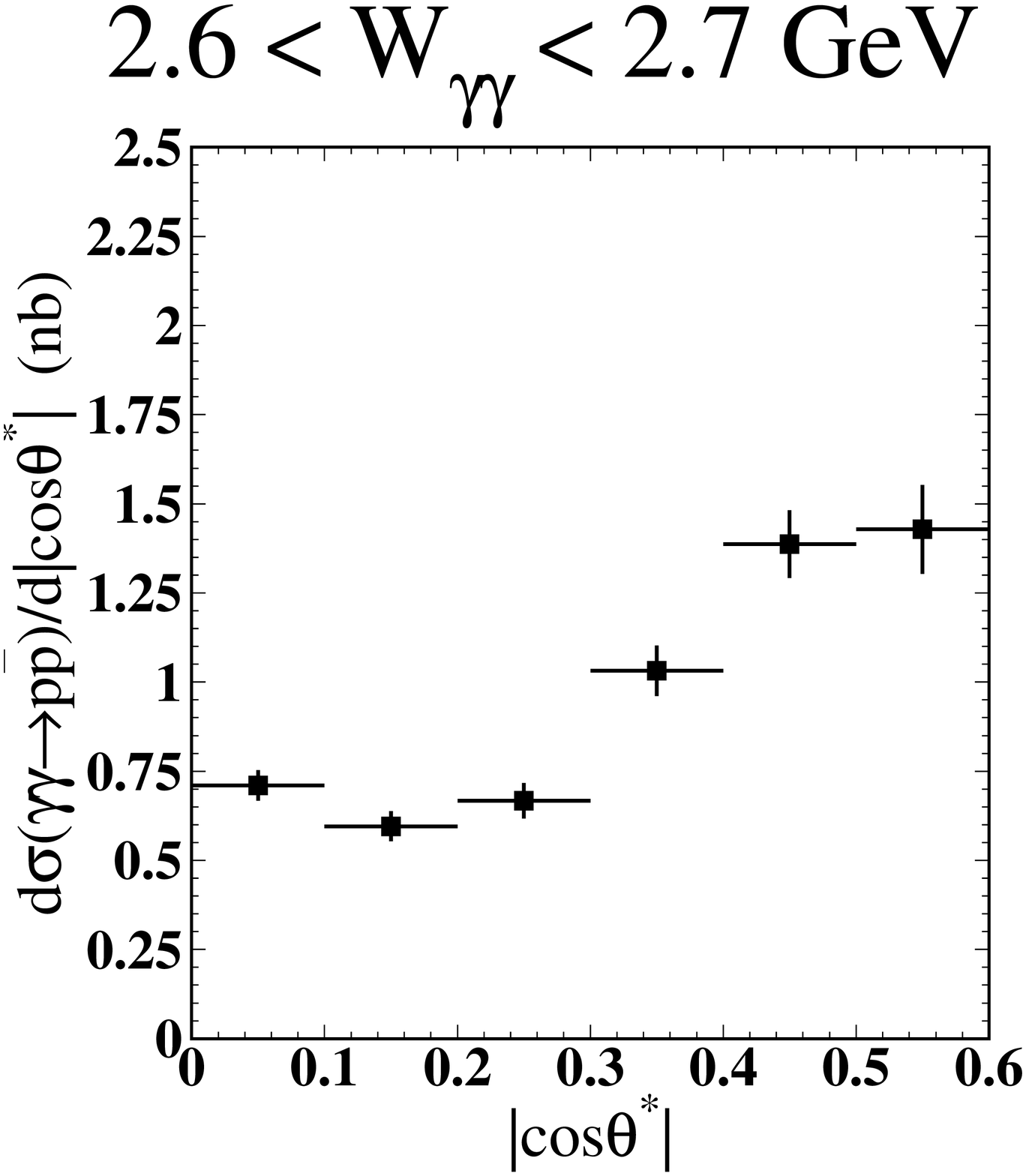}
\includegraphics[width=4.55cm,height=4.55cm]{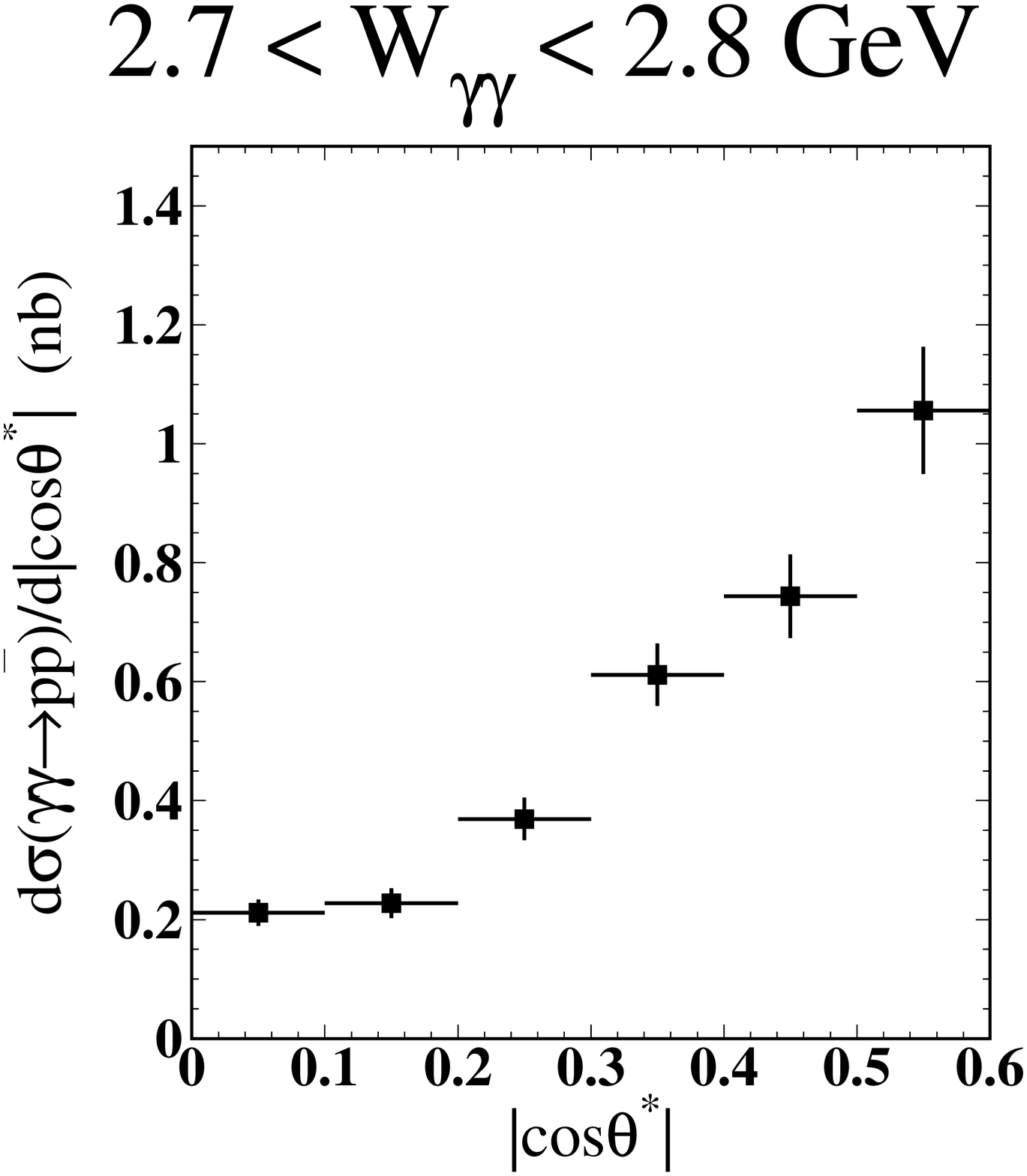}
\includegraphics[width=4.55cm,height=4.55cm]{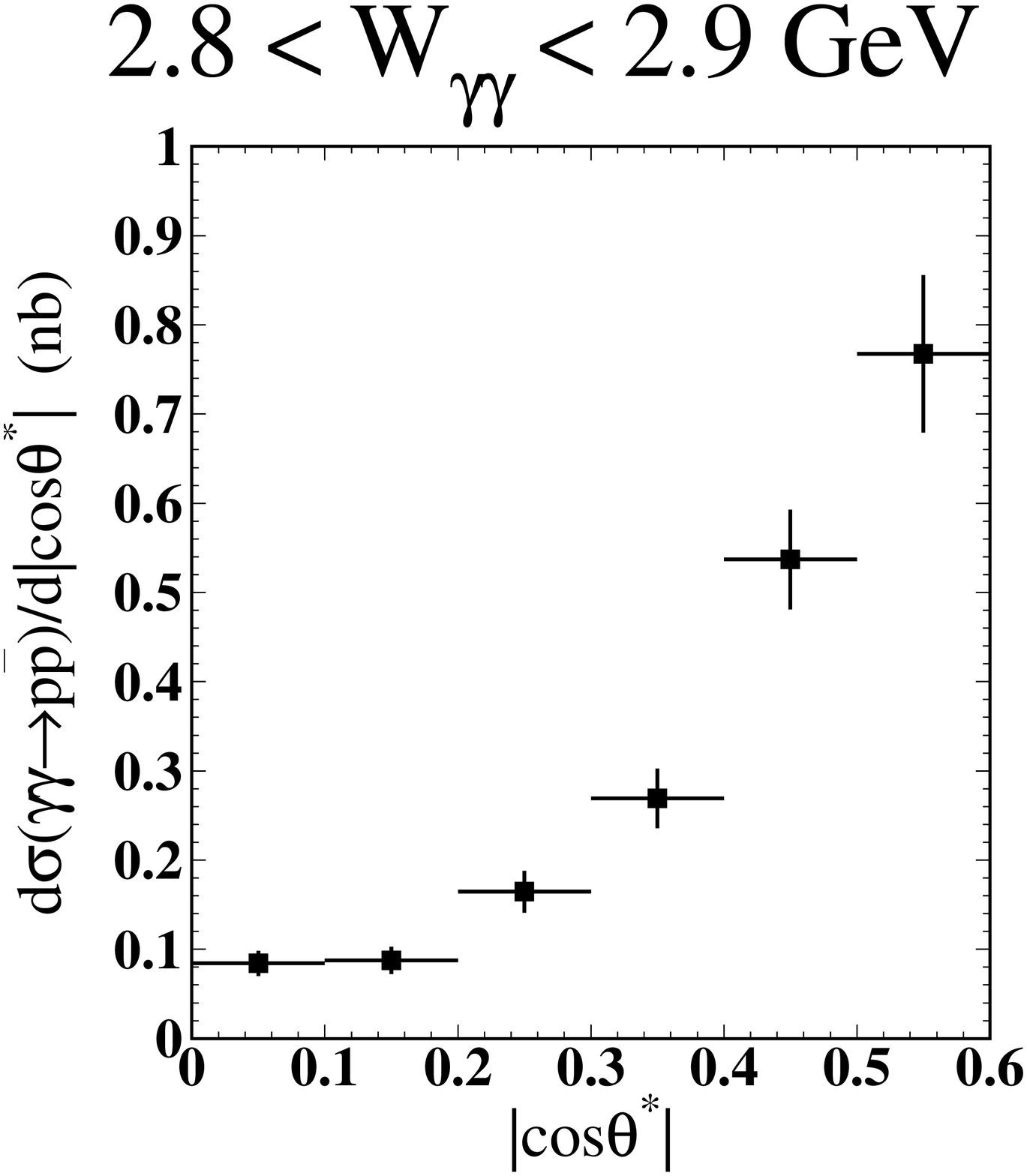}
\includegraphics[width=4.55cm,height=4.55cm]{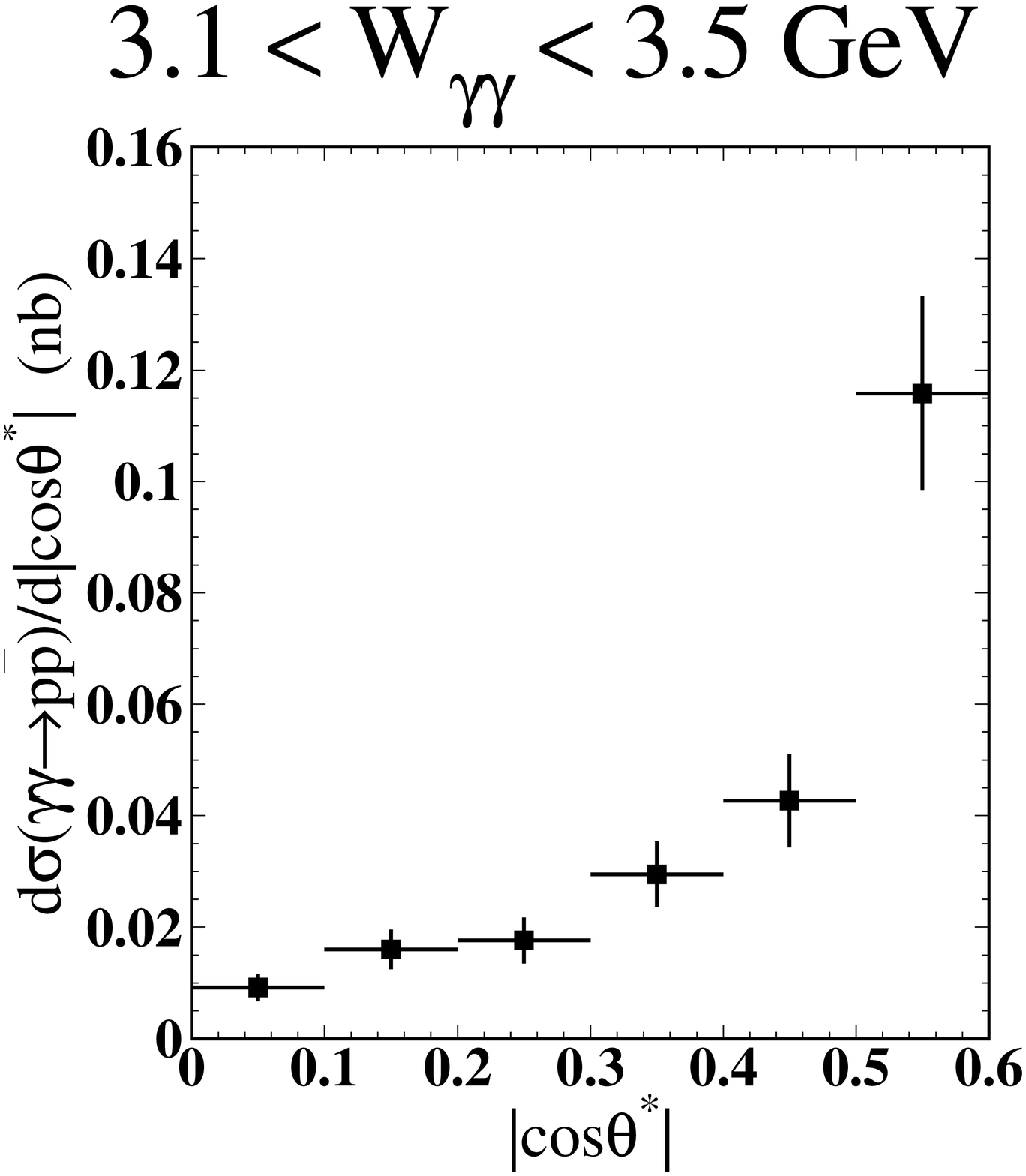}
\includegraphics[width=4.55cm,height=4.55cm]{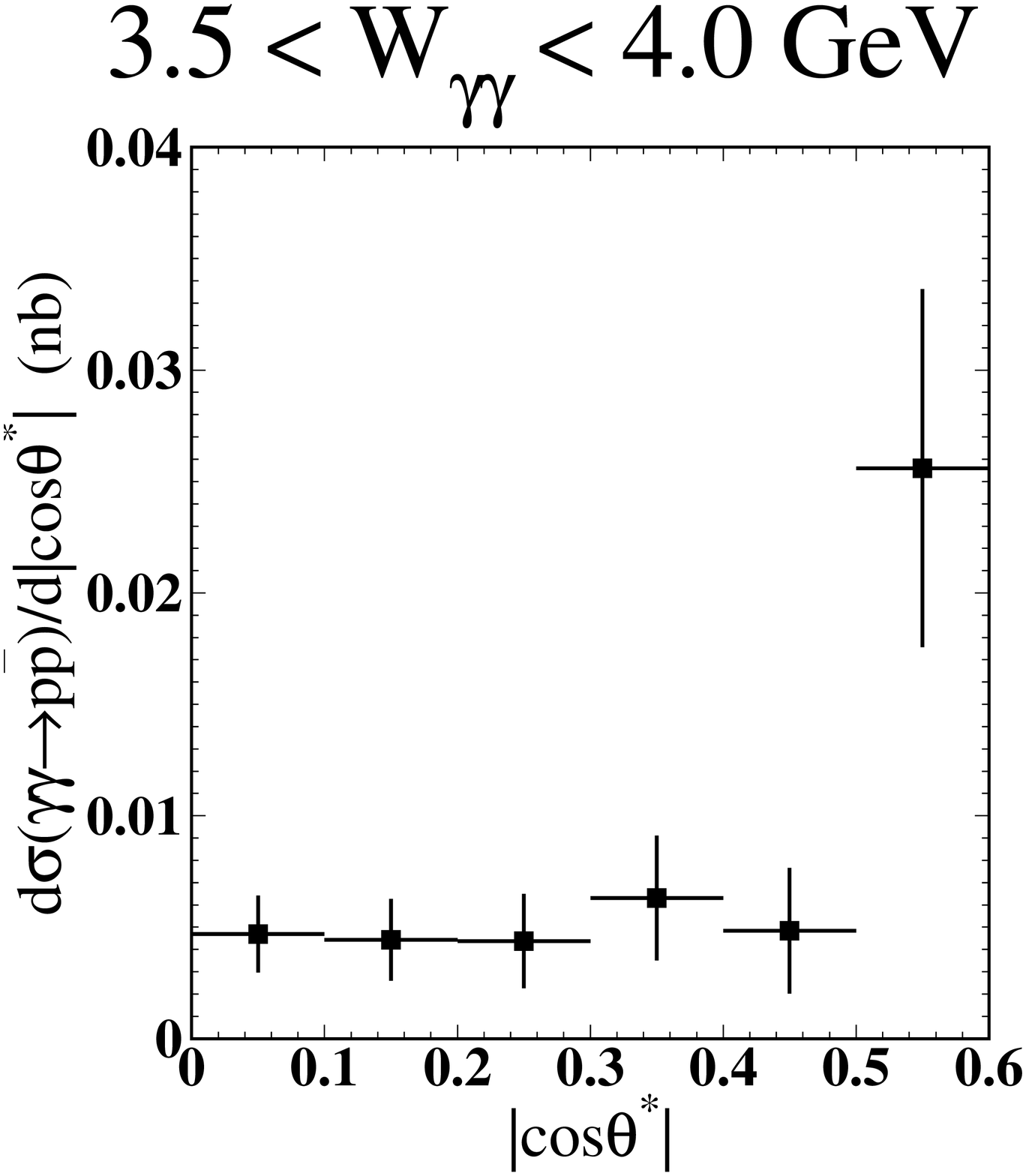}
\caption{Measured differential cross sections in 11 ranges of 
$W_{\gamma\gamma}$ as a function of $|\cos{\theta^*}|$. 
The error bars are statistical only.}
\end{center}
\end{figure}

Based on studies from Monte Carlo and data, 
residual backgrounds due to particle misidentification and non-exclusive events
are subtracted from the data. 
Corrections are based on 
$\Delta N$ multiplied by $(1-f)$ as shown in Eq.(2).
Complete details are given in Section 5, and the systematic errors are shown
in Table 3.
All measured cross sections and differential cross sections shown in this
paper have been corrected in this way. 
The excess caused by the $J/\psi$ background 
described at the end of Section 2
is estimated in each $|\cos{\theta^*}|$ bin separately and subtracted 
from $\Delta N$ before the other corrections above.
The systematic uncertainty due to this subtraction is $8\%$ for the
measured cross section in the 3.0 - 3.1 GeV
$W_{\gamma\gamma}$ bin, taking into account the 
fluctuation of the estimated number of $J/\psi$.

\begin{figure}
\begin{center}
\includegraphics[width=6.8cm,height=6.8cm]{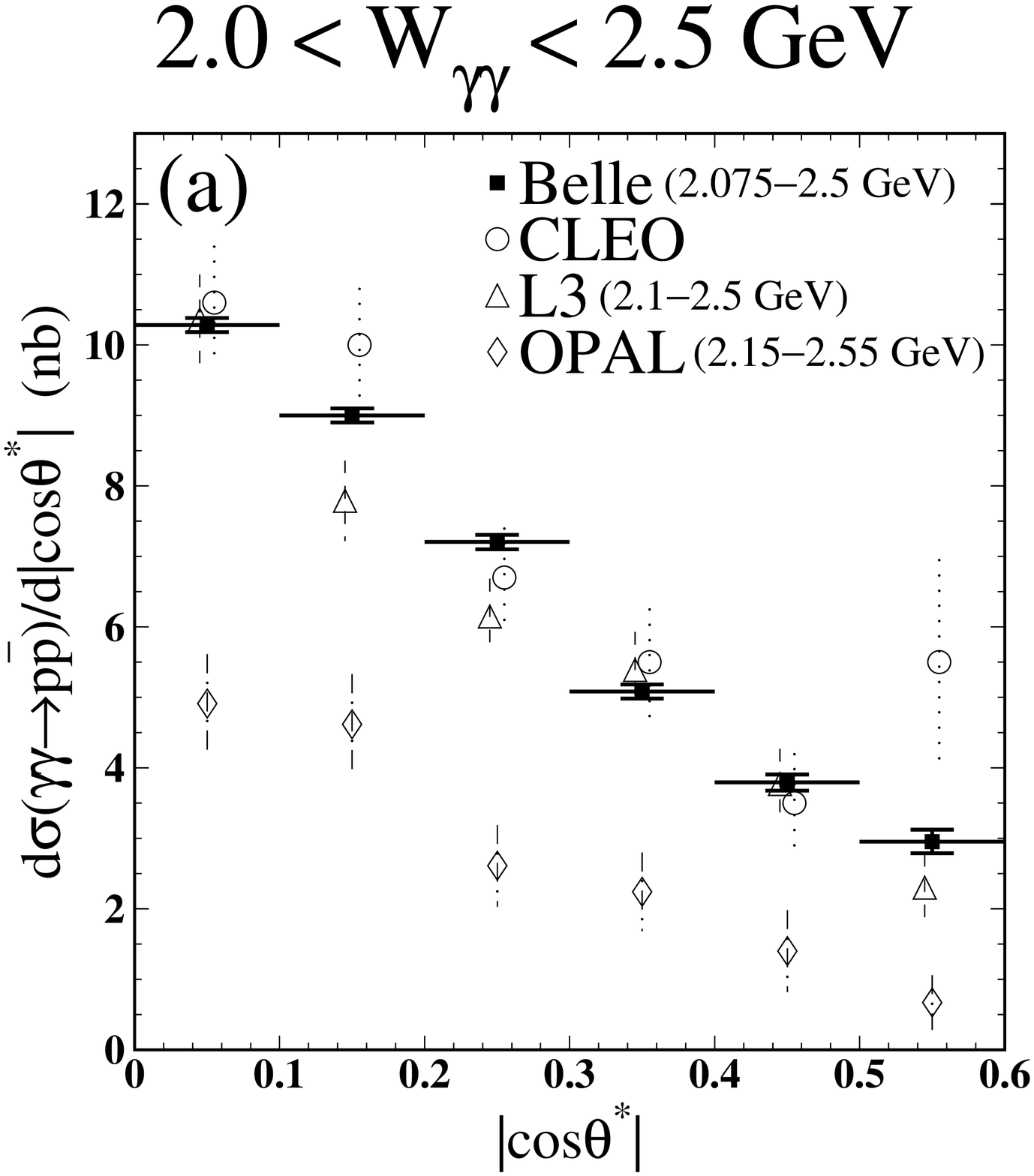}
\includegraphics[width=6.8cm,height=6.8cm]{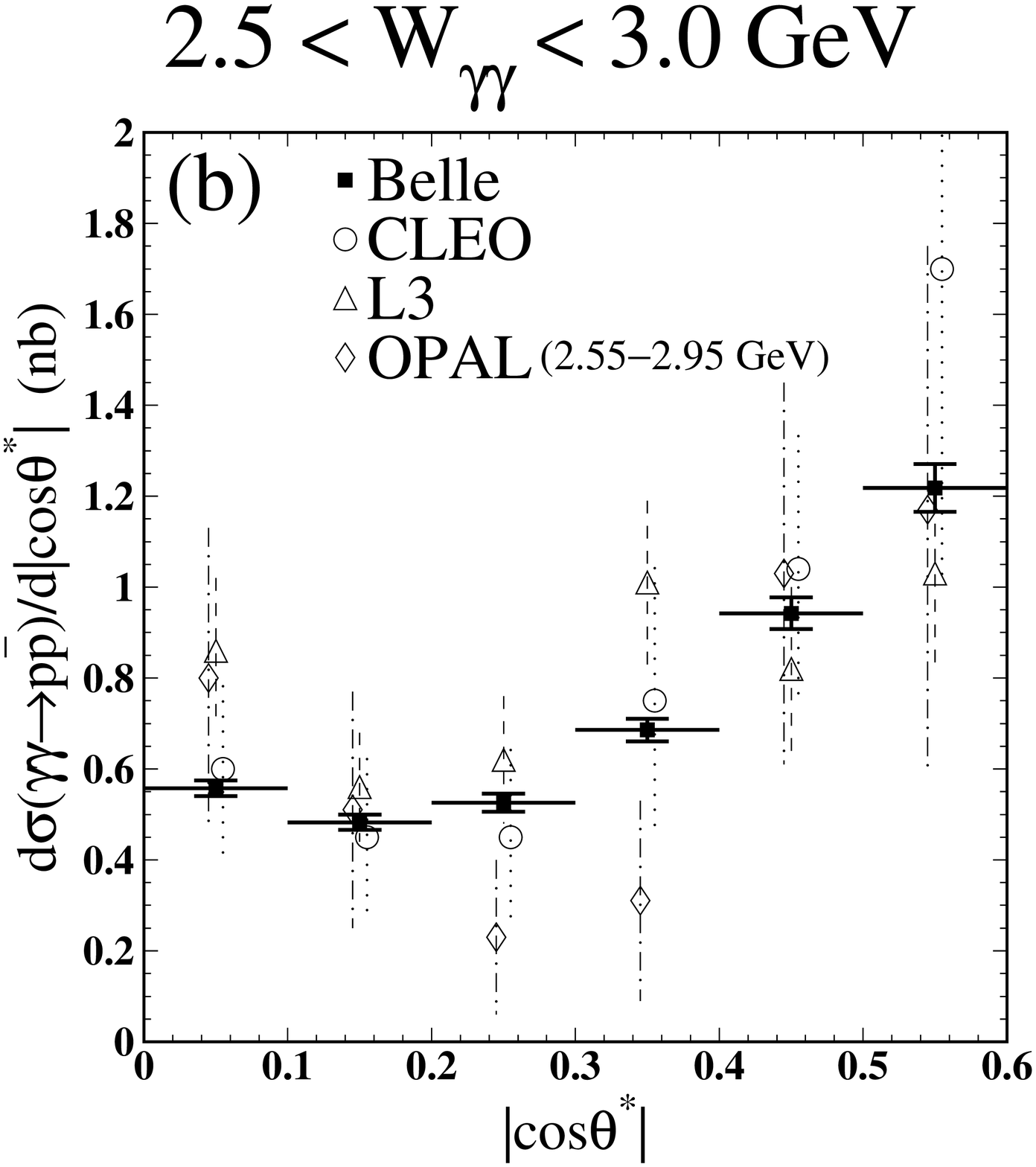}
\includegraphics[width=6.8cm,height=6.8cm]{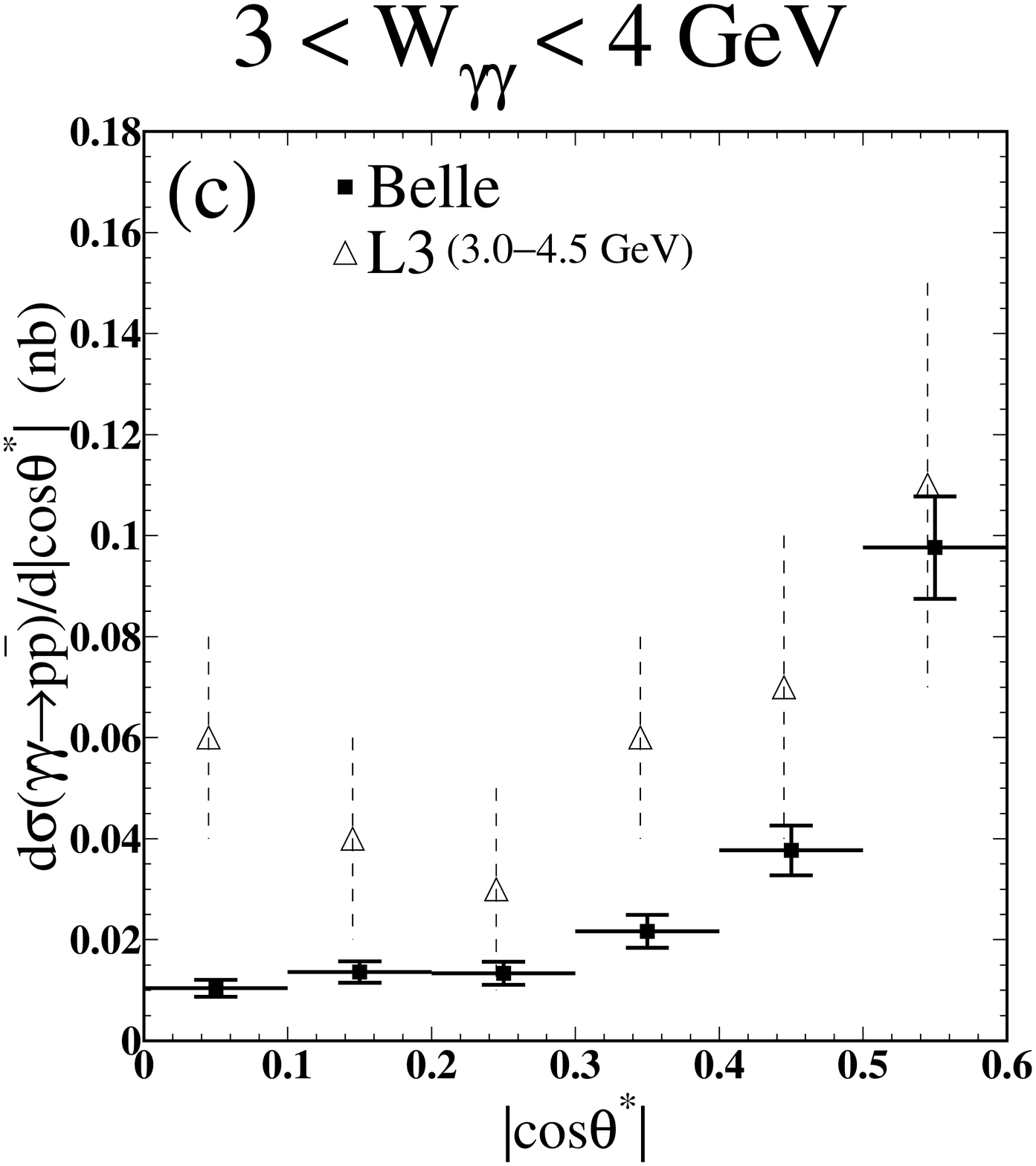}
\caption{Measured differential cross sections as a function of 
$|\cos{\theta^*}|$ for three ranges of $W_{\gamma\gamma}$. 
The error bars are statistical only.}
\end{center}
\end{figure}

\begin{table}
\begin{center}
{\footnotesize
\begin{tabular}{|c||c|c|c|}
\hline
 & $\frac{d\sigma_{\gamma\gamma\to p\overline{p}}}{d|\cos{\theta^*}|}$ (nb) &
   $\frac{d\sigma_{\gamma\gamma\to p\overline{p}}}{d|\cos{\theta^*}|}$ (nb) &
   $\frac{d\sigma_{\gamma\gamma\to p\overline{p}}}{d|\cos{\theta^*}|}$ (nb) \\
$|\cos{\theta^*}|$ & 
$W_{\gamma\gamma}=$2.075 - 2.5 GeV & $W_{\gamma\gamma}=$2.5 - 3.0 GeV & $W_{\gamma\gamma}=$3 - 4 GeV \\
\hline\hline
0.0 - 0.1 & 10.28 $\pm$ 0.10 $\pm$ 0.29 & 0.56 $\pm$ 0.02 $\pm$ 0.02 & 0.010 $\pm$ 0.002 $\pm$ 0.002 \\
0.1 - 0.2 & \hspace{0.2cm}9.00 $\pm$ 0.10 $\pm$ 0.26 & 0.48 $\pm$ 0.02 $\pm$ 0.02 & 0.014 $\pm$ 0.002 $\pm$ 0.001 \\
0.2 - 0.3 & \hspace{0.2cm}7.20 $\pm$ 0.10 $\pm$ 0.27 & 0.53 $\pm$ 0.02 $\pm$ 0.04 & 0.013 $\pm$ 0.002 $\pm$ 0.001 \\
0.3 - 0.4 & \hspace{0.2cm}5.08 $\pm$ 0.10 $\pm$ 0.19 & 0.69 $\pm$ 0.03 $\pm$ 0.05 & 0.022 $\pm$ 0.003 $\pm$ 0.002 \\
0.4 - 0.5 & \hspace{0.2cm}3.79 $\pm$ 0.12 $\pm$ 0.17 & 0.94 $\pm$ 0.04 $\pm$ 0.03 & 0.038 $\pm$ 0.005 $\pm$ 0.003 \\
0.5 - 0.6 & \hspace{0.2cm}2.96 $\pm$ 0.17 $\pm$ 0.13 & 1.22 $\pm$ 0.05 $\pm$ 0.04 & 0.098 $\pm$ 0.010 $\pm$ 0.007 \\
\hline
\end{tabular}
}
\caption{Measured differential cross sections 
versus $|\cos{\theta^*}|$ for
different $W_{\gamma\gamma}$ ranges.
The first error is statistical and the second is systematic.}
\end{center}
\end{table}

\begin{table}
\begin{center}
{\footnotesize
\begin{tabular}{lc}
\hline
Source & Systematic error (\%) \\
\hline
Integrated luminosity & 1.4\\
Luminosity function & 3 - 5\\
Trigger efficiency & 5\\
Particle identification efficiency & 1 - 6\\
Monte Carlo statistics & 1 - 3\\
Particle misidentification background subtraction & 0 - 1\\
Non-exclusive ($p\overline{p}\pi^0$) background subtraction & 2 - 12 \\
$|\Sigma p_t^*|$ effect and residual non-exclusive backgrounds & 2 - 3 \\
$J/\psi$ subtraction ($W_{\gamma\gamma}=3.0$ - 3.1 GeV) & 8\\
Possible backgrounds from radiative return & 1 \\
\hline
Total & 7 - 14 \\
\hline
\end{tabular}
}
\vspace{0.25cm}
\caption{Systematic errors for the measured cross sections of 
$\gamma\gamma\to p\overline{p}$. Some uncertainties are 
$W_{\gamma\gamma}$-dependent and shown as ranges.}
\end{center}
\end{table} 

The total $\eta_c$ yield, $N_{\eta_c}=156.9\pm33.3$, can be 
converted to the product of the two-photon width of $\eta_c$ and the 
branching fraction of $\eta_c\to p\overline{p}$:
$\Gamma_{\gamma\gamma}(\eta_c)\times B(\eta_c\to p\overline{p}) =
N_{\eta_c} m_{\eta_c}^2 / 
(4\pi^2\hspace{0.08cm} \varepsilon\hspace{0.08cm} L_{\rm int}\hspace{0.08cm}
\frac{dL_{\gamma\gamma}}{dW_{\gamma\gamma}}) =
7.20 \pm 1.53({\rm stat.}) ^{+0.67}_{-0.75}({\rm syst.})$ eV, using the 
luminosity function
$dL_{\gamma\gamma}/dW_{\gamma\gamma}$ determined at the energy of
the $\eta_c$ mass ($m_{\eta_c}$) and the efficiency $\varepsilon$ from 
Monte Carlo.  
For the systematic error, effects from the uncertainties 
of the continuum background shape and the signal width are taken into account, 
in addition to all other sources listed in Table 3. 
The above result gives $\Gamma_{\gamma\gamma}(\eta_c) =
5.5 \pm 1.2({\rm stat.}) ^{+0.5}_{-0.6}({\rm syst.}) \pm 1.7({\rm norm.})$ 
keV, where the last error
comes from the branching fraction $B(\eta_c\to p\overline{p})$ 
uncertainty \cite{PDG}. 
Since observations of the $\eta_c$ in the
$p\overline{p}$ channel are scarce and suffer from low statistics, 
current measurements for $B(\eta_c\to p\overline{p})$ available in Ref.~\cite{PDG} 
are not very consistent with each other.
Our result is the first measurement of 
$\Gamma_{\gamma\gamma}(\eta_c)\times B(\eta_c\to p\overline{p})$ 
in two-photon collisions and, together with the observation of the 
$\eta_c$ in $p\overline{p}$ collisions in its $\gamma\gamma$ decay 
mode~\cite{PP}, 
will help to decrease the errors on both the $\eta_c$ two-photon 
width and branching fraction.


\section{Corrections and major sources of systematic error}

The accuracy of the Monte Carlo trigger
efficiency has been checked from the two-track trigger, which requires at 
least two CDC tracks with an opening angle larger than $135^{\circ}$, 
two or more TOF hits as well as the ECL timing signal, using 
experimental events passing the high energy trigger based on a 
1~GeV threshold for an ECL 
total energy sum \cite{belle,ECLT}. 
The trigger efficiency depends 
on the average transverse momentum of the two tracks in the laboratory 
frame, $\overline{p}_t\cong p_t^{\gamma\gamma}\equiv
[(W_{\gamma\gamma}/2)^2-m_p^2]^{1/2} \cdot
(1-|\cos{\theta^*}|^2)^{1/2}$, where the latter is
the transverse momentum of $p(\overline{p})$ in 
the $\gamma\gamma$ c.m. frame. We determine the trigger efficiency
as a function of $p_t^{\gamma\gamma}$, since
each of the two-dimensional bins in $W_{\gamma\gamma}$ 
and $|\cos{\theta^*}|$, where the number of events is measured,
is associated with a $p_t^{\gamma\gamma}$ value using the relation above.
From the data, 
the trigger efficiency is
$0.83\pm0.02$ at $p_t^{\gamma\gamma}=0.55$ GeV/$c$ and 
$0.95\pm0.05$ at $p_t^{\gamma\gamma}=0.95$ GeV/$c$. 
Corrections for the Monte Carlo trigger efficiency are implemented 
according to the data, with a systematic error within $5\%$. 

The accuracy of the PID efficiencies has been checked 
by comparing Monte Carlo estimates to those based on data. The 
efficiency associated with each of the four PID conditions (Section 2) 
is studied,
using events passing all selection criteria except
the condition in question. 
The overall PID efficiency is $\sim 92\%$, $\sim 88\%$ and 
down to $\sim 78\%$ at $W_{\gamma\gamma}=2$, 3 and 4 GeV respectively,
with a systematic error 
less than $6\%$ in the whole $W_{\gamma\gamma}$ range.
The fake rate is $\sim 0.01\%$ - $0.3\%$ 
for $W_{\gamma\gamma}=3$ - 4 GeV, respectively.   

Monte Carlo studies indicate that the PID requirements are very 
efficient in rejecting electrons and other relativistic 
particles, so that events from  
$\gamma\gamma\to \pi^+\pi^-,\mu^+\mu^-$ and $e^+e^-$ do 
not survive the selection, leaving those from 
$\gamma\gamma\to K^+K^-$ as the main residual background.
From Monte Carlo simulation and the measured cross sections for 
$\gamma\gamma\to K^+K^-$ \cite{PI2,UH1},
the fraction of data that can be 
attributed to residual $K^+K^-$ background, $f_m$,
is evaluated. Based on Monte Carlo studies,
the dependence of $f_m$ on $|\cos{\theta^*}|$ is negligible and
$f_m(W_{\gamma\gamma})=0.8\pm 0.3\%$, $3.2\pm 0.5\%$ and $7.7\pm 0.8\%$
at $W_{\gamma\gamma}=3.2$, 3.6 and 4.0 GeV, respectively.
For $W_{\gamma\gamma}<3.0$ GeV, $f_m$ is negligible.
The values of $f_m$  have been checked in the data, using events 
passing all selection criteria except that on the normalized
likelihood, $\lambda_p>0.8$.
The number of signal events that would pass all selection criteria is 
estimated from the $\Delta T$ 
distribution of one of the two tracks, after requiring the other 
to satisfy $\lambda_p>0.8$.
The values of $f_m$ inferred in this way   
are in good agreement with those above.   
The contribution from this source of background 
is subtracted from the data, 
using the expected $f_m$ from 
the Monte Carlo studies.  
The systematic uncertainty due to this source is $\sim 1\%$ or less
in the whole $W_{\gamma\gamma}$ range.  

Possible non-exclusive backgrounds ($p\overline{p}X$),
most of them from $\gamma\gamma\to p\overline{p}\pi^0$ events,
have been searched for in the data. 
Monte Carlo studies show that a high purity sample of such 
background can be derived from events with larger $|\Sigma p_t^*|$
and smaller $|\Sigma p_t^*(p\overline{p}\pi^0)|$,
the latter being the transverse momentum balance of the three particles. 
By comparing Monte Carlo and data distributions of these parameters
we obtain the fraction of the data attributed to this background type,
$f_n$. We find that the dependence of $f_n$ on $|\cos{\theta^*}|$ is negligible,  
and it ranges from $5\pm2\%$ to $17\pm8\%$ for
$W_{\gamma\gamma}$ from 2 to 4 GeV, respectively. Corrections are made 
using the $f_n(W_{\gamma\gamma})$ obtained above, 
and in total $7\pm1\%$ of the selected data are subtracted. 
The systematic error from this source is 2 - $12\%$ 
for $W_{\gamma\gamma}$ from 2 to 4 GeV, respectively.  
The fraction of the data attributed to 
$\gamma\gamma\to p\overline{p}\pi^0$ events
is also obtained as a function of $|\Sigma p_t^*|$.
Before the correction,
a comparison of the $|\Sigma p_t^*|$ distribution between data and Monte Carlo 
exhibits a total difference of $\sim 9\%$ between the 
two samples, while it is reduced to less than $3\%$ 
for any $W_{\gamma\gamma}$ range after the correction (Fig.~7).
Since the residual excess in the $|\Sigma p_t^*|$ distribution
could be attributed to  
residual non-exclusive backgrounds and a broader nature of the signal 
distribution than the Monte Carlo, the systematic uncertainty due to
other possible non-exclusive backgrounds is limited to $3\%$ after the
correction.  
    
\begin{figure}[h]
\begin{center}
\includegraphics[width=11.2cm,height=11.2cm]{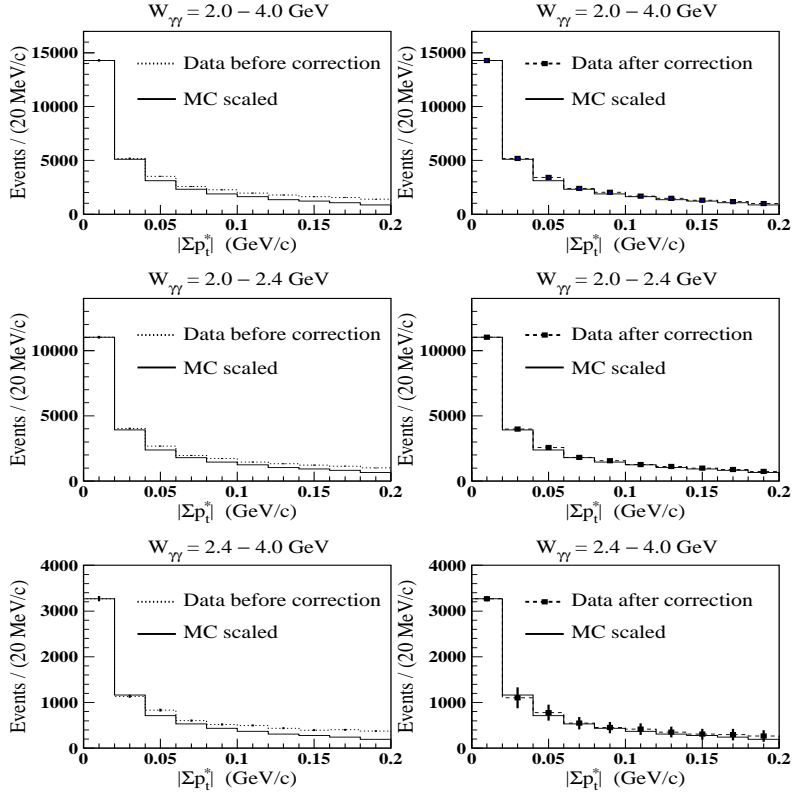}
\caption{$|\Sigma p_t^*|$ distributions for the data
before (left column) and after (right column) the subtraction of 
residual non-exclusive backgrounds ($\gamma\gamma\to p\overline{p}\pi^0$).
The Monte Carlo distributions are scaled with the 
first bin normalized to the data.}
\end{center}
\end{figure} 
   

\section{Theoretical approaches}   

\begin{figure}
\begin{center}
\includegraphics[width=6.8cm,height=6.8cm]{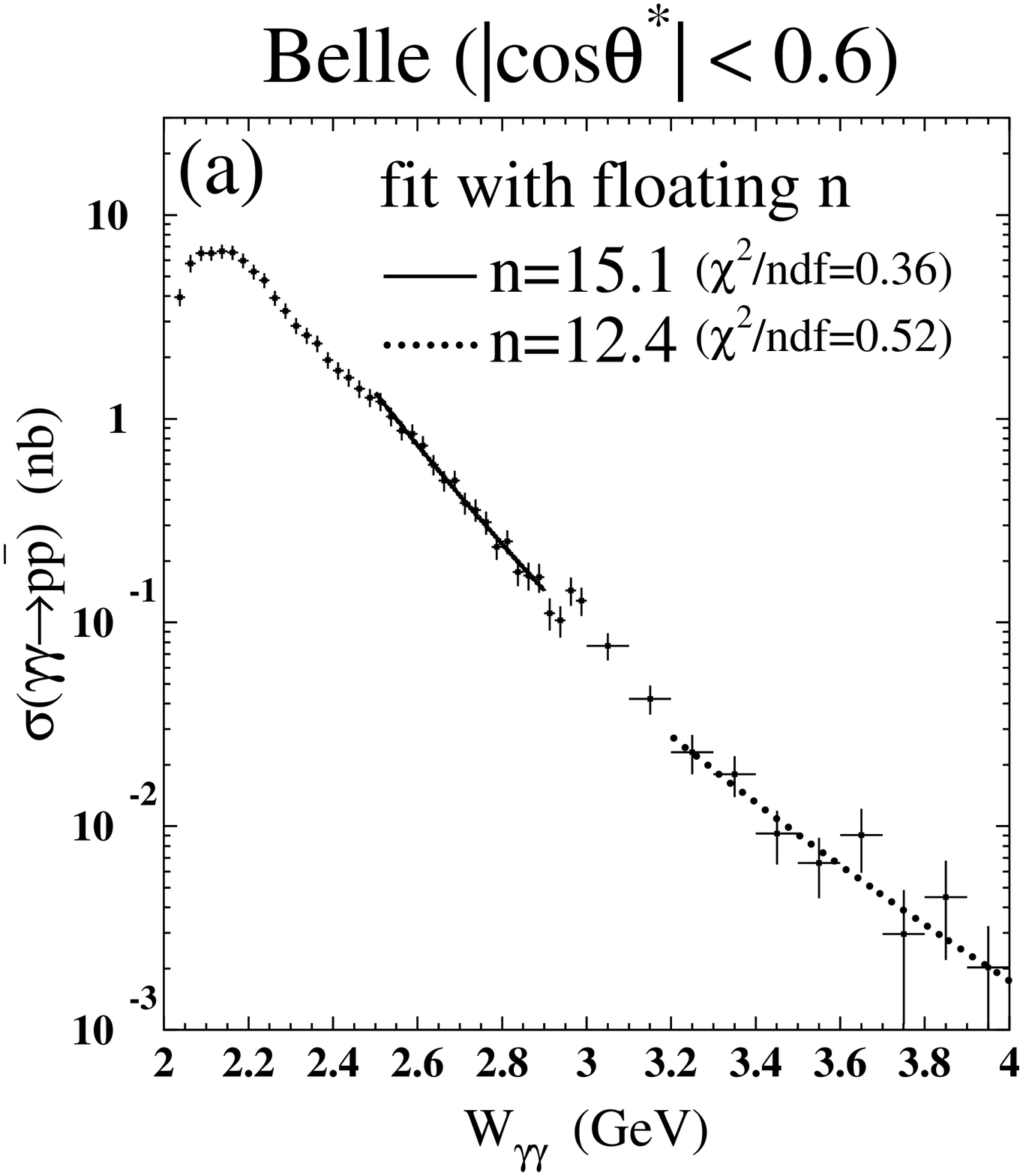}
\includegraphics[width=6.8cm,height=6.8cm]{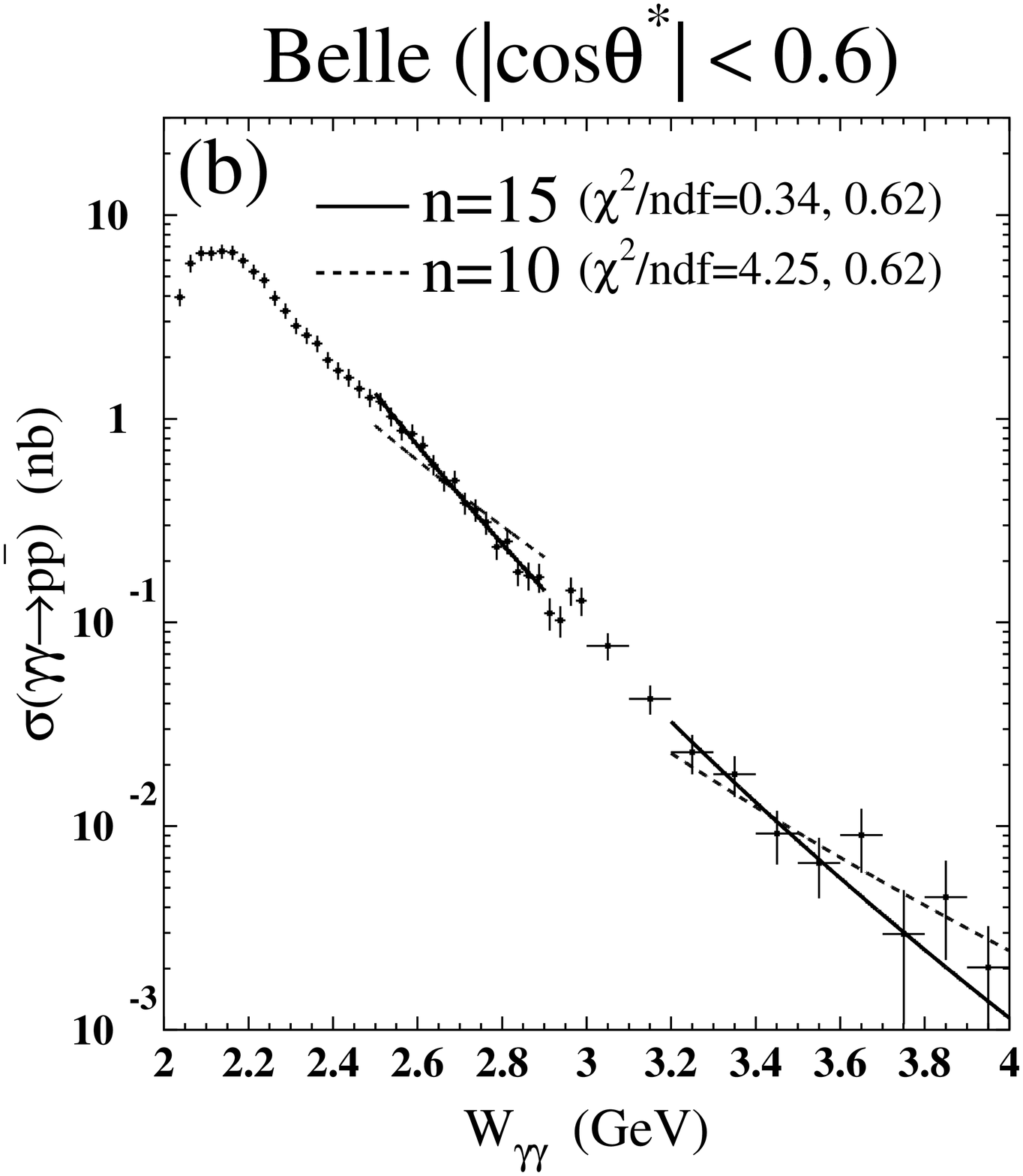}
\caption{Separate fits of $\sigma\propto W_{\gamma\gamma}^{-n}$  
to the data in the range of $W_{\gamma\gamma}=2.5$ - 2.9 GeV and
3.2 - 4.0 GeV, with (a) $n$ floating; (b) $n=10$ and $n=15$. 
The error bars include statistical and systematic errors.
The $\chi^2/{\rm ndf}$ values for each fit are indicated in the figure.}  
\end{center}
\end{figure} 

From the asymptotic QCD prediction of Eq.(1) 
and after integration over $\cos{\theta^*}$, 
the cross section for $\gamma\gamma\to p\overline{p}$ 
is proportional to $W_{\gamma\gamma}^{-10}$
for asymptotically large $W_{\gamma\gamma}$.
All models based on this 
framework behave asymptotically as 
$\sigma\propto W_{\gamma\gamma}^{-10}$. 
For the diquark scenario,  
two curves are provided \cite{CF5}: from the complete diquark model and from
the same model with only helicity conserved amplitudes, where $p$ and 
$\overline{p}$
are in opposite helicity states. 
The scale of the diquark predictions matches the data
for $W_{\gamma\gamma}=2.5$ - 3.0 GeV, 
but the deviation becomes larger as $W_{\gamma\gamma}$ increases (Fig.~4).
At higher energies, the data fall below the diquark 
predictions and exhibit a gradual approach to the three-quark model
predictions \cite{FA}. At medium energies between 2.5 and 4.0 GeV, 
a steeper fall of the total cross section in $W_{\gamma\gamma}$ is 
observed. 

If we fit the data with a power law  
$\sigma\propto W_{\gamma\gamma}^{-n}$ with $n$ floating (Fig.~8(a)),
taking into account both statistical and systematic 
uncertainties as well as possible correlations between the latter,  
we obtain $n=15.1^{+0.8}_{-1.1}$ and $12.4^{+2.4}_{-2.3}$  
in the range of 
$W_{\gamma\gamma}=2.5$ - 2.9 GeV and 
3.2 - 4.0 GeV, respectively 
(the 
charmonium region between 2.9 and 3.2 GeV is excluded).
For completeness, we also show in Fig.~8(b) 
the results of the fits with $n$ fixed at 10 and 15.
Although for both ranges a good fit to the data can be obtained at $n=15$, 
a smaller power, $n=10$, describes the data above 3.2 GeV reasonably well.
This may imply that lower power terms become dominant at higher enegies,
which is an indication for the transition to the asymptotic predictions.

The angular differential cross section in $|\cos{\theta^*}|$ 
is another observable most important to 
the study of the nature of the interactions involved in the process
$\gamma\gamma\to p\overline{p}$.
All existing models based on the constituent scattering picture
\cite{FA,FA2,MA,KP,CF5,HB}, as expected, predict an ascending trend,
which is in agreement with the data for $W_{\gamma\gamma}>2.5$ GeV.  
This is due to the factor $1/\sqrt{tu}\propto 1/\sqrt{1-\cos^2{\theta^*}}$
contained in the hard scattering amplitudes.
The same trend is obtained from naive QED \cite{QED} estimates:  
$d\sigma/d|\cos{\theta^*}|\propto 
(1+\cos^2{\theta^*})/(1-\cos^2{\theta^*})$, in the massless limit. 
A simplified picture with diquarks would follow the naive QED 
expectation above \cite{OP,MA}, if all quark masses are neglected and 
only scalar diquarks are considered.
In Fig.~9 we plot 
the differential cross section normalized to that averaged within
$|\cos{\theta^*}|<0.3$, and compare various predictions 
to the data. We observe that the data 
rise more sharply in $|\cos{\theta^*}|$ at
higher energy (see also Fig.~5). In comparison, all current models
predict a flatter trend in $|\cos{\theta^*}|$.     

\begin{figure}[h]
\begin{center}
\includegraphics[width=6.8cm,height=6.8cm]{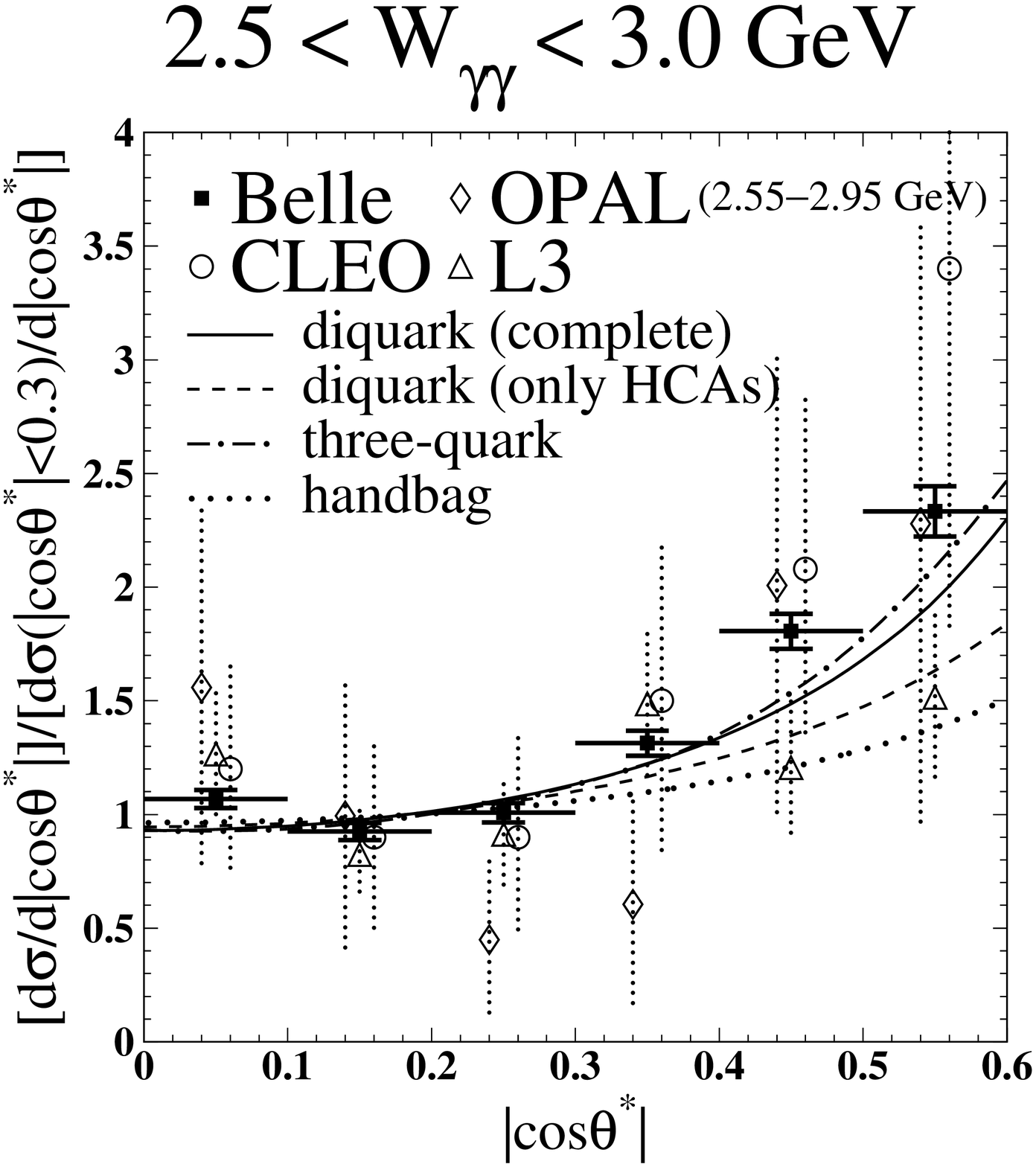}
\includegraphics[width=6.8cm,height=6.8cm]{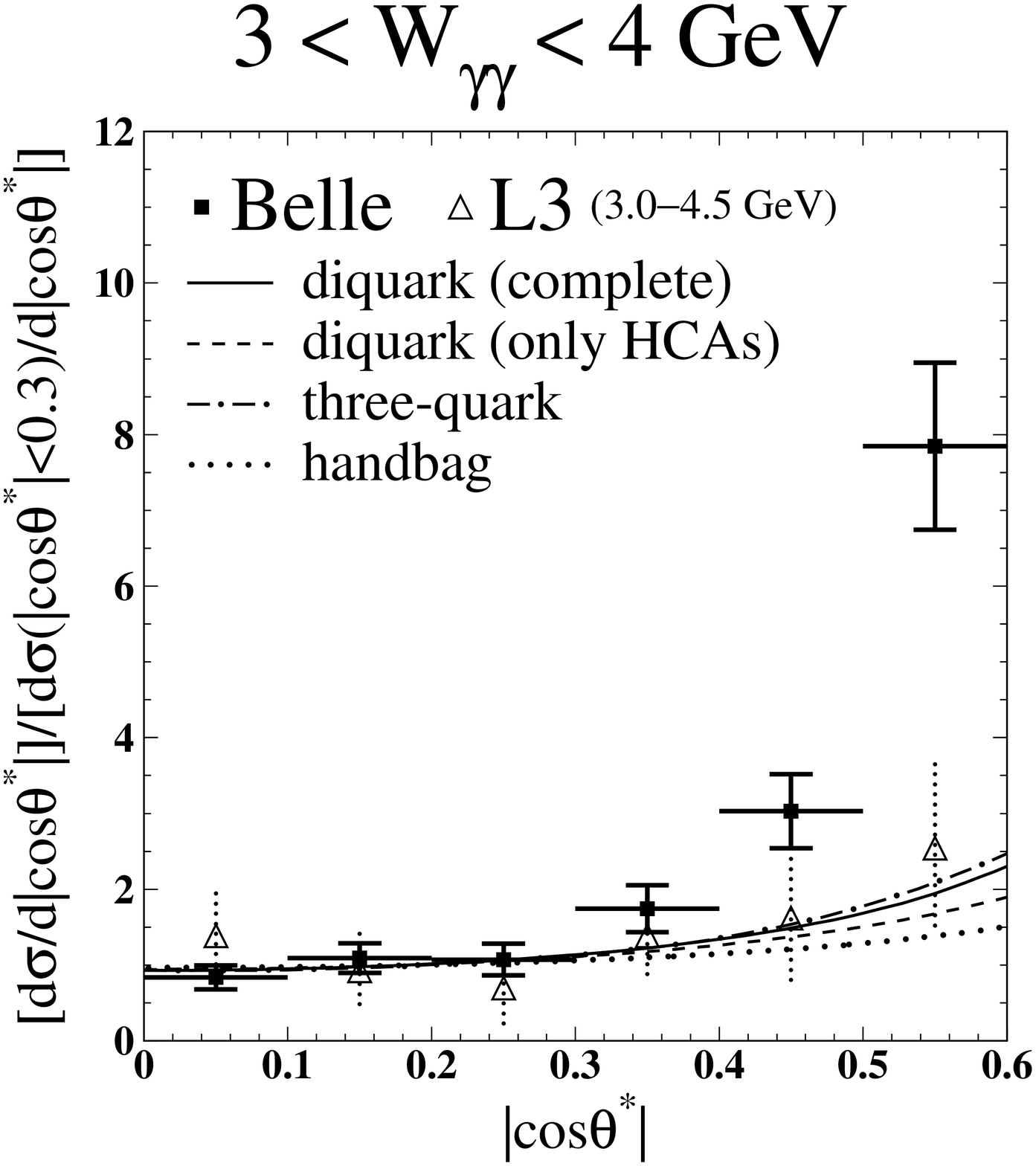}
\caption{Differential cross section as a function of $|\cos{\theta^*}|$, 
normalized to 
that averaged within $|\cos{\theta^*}|<0.3$, 
for the two higher ranges of $W_{\gamma\gamma}$. The error bars are  
statistical only.
Theoretical predictions are from \cite{CF5} (diquark), 
\cite{FA2} (three-quark) and \cite{HB} (handbag).}
\end{center}
\end{figure}

The deviation of the leading term QCD calculations~\cite{FA,FA2} 
from the data at $W_{\gamma\gamma}=2.5$ - 4.0 GeV
implies that power corrections are still significant 
at these intermediate energies. It is not surprising since the very
threshold of $p\overline{p}$ production corresponds to
$W_{\gamma\gamma} \sim 2$~GeV. 
However, the diquark and handbag 
models~\cite{MA,KP,CF5,HB} 
were developed in order to
describe the intermediate 
energy region at the price of introducing model form factors, {\it etc.}.
The disagreement of the data at  $W_{\gamma\gamma}=2.5$ - 4.0 GeV
with their predictions (see Fig.~4 and 9) obviously
necessitates their improvement.  

The descending trend of the differential cross section in $|\cos{\theta^*}|$
observed at low energies ($W_{\gamma\gamma}<2.5$ GeV) cannot be 
understood within the hard scattering picture (Fig.~6(a)).
In a recent study based on non-perturbative QCD sum rules \cite{HL2}, 
this trend was proposed as a general feature for 
hadron pair production
from two-photon collisions. The behavior
is very natural if low partial waves are involved. In Ref.~\cite{OD} 
it was shown that even a simple model based on pole- and 
resonance-dynamics can reproduce this behavior.          


\section{Conclusion}

Using the Belle detector at the high-luminosity KEKB collider,  
the cross sections for $\gamma\gamma\to p\overline{p}$
have been measured for $W_{\gamma\gamma}$ from 2.025 to 4.0 GeV and
$|\cos{\theta^*}|<0.6$, with
systematic uncertainties ranging from 7\% to 14\%. 
These results represent a great improvement in precision
compared to all previous measurements and allow more accurate tests of  
various theoretical models. We also observed the
production of $\gamma\gamma\to\eta_c\to p\overline{p}$ and determined  
the product of the two-photon width of the $\eta_c$ and 
its branching ratio to $p\overline{p}$. 

Fitting to a power law $\sigma\propto W_{\gamma\gamma}^{-n}$ 
shows that the best fit
value of $n$ decreases as energy increases, and $n=10$ cannot be rejected at
energies above 3.2 GeV, implying the gradual transition 
to the expectation from asymptotic predictions. 
The ascending trend for the differential 
cross section in 
$|\cos{\theta^*}|$ predicted by the hard scattering picture 
is in agreement with the data for $W_{\gamma\gamma}>2.5$ GeV; however,
the data 
rise more sharply in $|\cos{\theta^*}|$ as $W_{\gamma\gamma}$ increases.
The descending trend in $|\cos{\theta^*}|$ at lower energies  
$W_{\gamma\gamma} < 2.5$ GeV can be reproduced by non-perturbative 
approaches~\cite{OD}.  
The descending trend of the differential cross section in $|\cos{\theta^*}|$
changes to an ascending one with the increase of energy, 
which could be an indication for the 
transition from a soft resonance regime to the beginning of a hard regime.
Existing models suggested for the intermediate energies~\cite{MA,KP,CF5,HB}
can not provide satisfactory description of the
observed energy and angular dependence in the studied energy range.

\section*{Acknowledgements}

We thank the KEKB group for the excellent operation of the
accelerator, the KEK cryogenics group for the efficient
operation of the solenoid, and the KEK computer group and
the NII for valuable computing and Super-SINET network
support.
We thank C.~F.~Berger, V.~L.~Chernyak, H.-N.~Li and K.~Odagiri
for fruitful discussions.  
We acknowledge support from MEXT and JSPS (Japan);
ARC and DEST (Australia); NSFC (contract No.~10175071,
China); DST (India); the BK21 program of MOEHRD and the CHEP
SRC program of KOSEF (Korea); KBN (contract No.~2P03B 01324,
Poland); MIST (Russia); MESS (Slovenia); SNSF (Switzerland); 
NSC (under Grant No.~NSC-92-2112-M-008-028)
and MOE (Taiwan); and DOE (USA). 


\end{document}